\documentclass[a4paper]{article}
\usepackage{natbib}
\usepackage{graphicx}
\usepackage{a4wide}
\usepackage[utf8]{inputenc}
\usepackage[english]{babel}
\newcommand{\mitbf}[1]{
  \hbox{\mathversion{bold}$#1$}}
\newcommand{\tstep}{\Delta t}

\newcommand{\bdisc}{\mathbf{b}}
\newcommand{\udisc}{\mathbf{u}}

\newcommand{\fudisc}{\mathbf{f}}
\newcommand{\fbdisc}{\mathbf{f}}
\newcommand{\ub}[1]{\left[%
                    \begin{array}{l}%
                    \udisc_{#1}\\%
                    \bdisc_{#1}%
                    \end{array}%
                    \right]}
\newcommand{\dub}[1]{\left[%
                    \begin{array}{l}%
                    \delta \udisc_{#1}\\%
                    \delta \bdisc_{#1}%
                    \end{array}%
                    \right]}
\newcommand{\fub}[1]{\left[%
                    \begin{array}{l}%
                    \fudisc_{u,#1}\\%
                    \fbdisc_{b,#1}%
                    \end{array}%
                    \right]} 
\newcommand{\ltwo}[1]{\left\| #1 \right\|}
\newcommand{\helmu}{\mathbf{\sf H}_u}
\newcommand{\helmb}{\mathbf{\sf H}_b}
\newcommand{\massm}{\mathbf{\sf M}}
\newcommand{\massi}{\mathbf{\sf I}}
\newcommand{\massio}{\mathbf{\sf I}^o}
\newcommand{\stiff}{\mathbf{\sf K}}
\newcommand{\deriv}{\mathbf{\sf D}}
\newcommand{\mata}{\mathbf{\sf A}}
\newcommand{\matb}{\mathbf{\sf B}}
\newcommand{\matc}{\mathbf{\sf C}}
\newcommand{\matconst}{{C}}
\newcommand{\becm}{\mathbf{B}}
\newcommand{\mate}{\mathbf{\sf E}}
\newcommand{\eb}{e^b}
\newcommand{\eu}{e^u}
\newcommand{\hspa}{H^{\mbox{S}}}
\newcommand{\statev}{\mathbf{x}}
\newcommand{\mparam}{\mathbf{p}}
\newcommand{\noise}{\mitbf{\epsilon}}
\newcommand{\descent}{\mathbf{d}}
\newcommand{\adjoint}{\mathbf{a}}
\newcommand{\observ}{\mathbf{y}^o}
\newcommand{\obscor}{\mathbf{R}}
\newcommand{\had}{\odot}
\newcommand{\grd}{\mitbf{\nabla}}
\newcommand{\nsta}{n^{\mbox{S}}}

\newcommand{\ecov}{\obscor}
\newcommand{\eqref}[1]{(\ref{#1})}
\begin{document}
\title{A case for variational geomagnetic data assimilation: insights from a one-dimensional,
       nonlinear, and sparsely observed MHD system \footnote{Article published in Nonlinear
 Processes in Geophysics, {\bf 14}, 163--180, 2007. The paper can be freely downloaded from
the journal webpage {\tt http://www.copernicus.org/EGU/npg/npg.html}}}
\author{A. Fournier, C. Eymin and T. Alboussi\`ere \footnote{Laboratoire de G\'eophysique Interne et Tectonophysique, Universit\'e Joseph-Fourier,
       BP 53, 38041 Grenoble cedex 9, France. Correspondence: alexandre.fournier@ujf-grenoble.fr}}
\maketitle

\begin{abstract}
Secular variations of the geomagnetic field have been measured with a
continuously improving accuracy during the last few hundred years,
culminating nowadays with satellite data. It is however well known that the
dynamics of the magnetic field is linked to that of the velocity field in
the core and any attempt to model secular variations will involve a
coupled dynamical system for magnetic field and core velocity.
Unfortunately, there is no direct observation of the velocity.
Independently of the exact nature of the above-mentioned coupled system --
some version being currently under construction -- the question is debated
in this paper whether good knowledge of the magnetic field can be
translated into good knowledge of core dynamics. Furthermore, what
will be the impact of the most recent and precise geomagnetic data on our
knowledge of the geomagnetic field of the past and future? These questions
are cast into the language of variational data assimilation, while the dynamical system
considered in this paper consists in a set of two oversimplified
one-dimensional equations for magnetic and velocity fields. This toy model
retains important features inherited from the induction and Navier-Stokes
equations: non-linear magnetic and momentum terms are present and its
linear response to small disturbances contains Alfv\'en waves. It is
concluded that variational data assimilation is indeed appropriate in principle, even
though the velocity field remains hidden at all times; it allows us to
recover the entire evolution of both fields from partial and irregularly
distributed information on the magnetic field. This work constitutes a
first step on the way toward the reassimilation of historical geomagnetic
data and geomagnetic forecast.
\end{abstract}

\section{Introduction}
The magnetic observation of the earth with satellites has now matured
to a point where continuous measurements of the field are available 
from 1999 onwards, thanks to the Oersted, SAC-C, and CHAMP missions
\cite[e.g.][and references therein]{orsted2000,pomme}. 
In conjunction with ground-based measurements, such data have been
used to produce a main field model of remarkable accuracy, in particular
concerning the geomagnetic secular variation (GSV)\citep{chaos}. Let us stress
that we are concerned in this paper with recent changes in the
earth's magnetic field, occurring over time scales on the order
of decades to centuries. This time scale is nothing compared to
the age of the earth's dynamo ($>3$ Gyr), or the average period at
which the dynamo reverses its polarity (a few hundreds of kyr, see
for instance \cite{mef96}), or even the magnetic diffusion time scale
in earth's core, on the order of $10$ kyr \citep[e.g.][]{bpc96}. 
 It is, however, over this minuscule time
window that the magnetic field and its changes are by far 
best documented \citep[e.g.][]{1989bgj}. 

Downward-projecting the surface magnetic field at the core-mantle
boundary, and applying the continuity of the normal component
of the field across this boundary, one obtains a map of this particular  
component at the top of the core. The catalog of these maps
at different epochs constitutes most of the data we have at hand 
to estimate the core state. Until now, this data has been 
exploited within a kinematic framework \citep{rs1965,backus1968}:  
the normal component of the magnetic field is a passive tracer, 
the variations of which are used to infer the velocity that
transports it \citep[e.g.][]{lemouel1984,bloxham1989}. 
 For the purpose of modeling the core field and interpreting its temporal
variations not only in terms of core kinematics,  
but more importantly in terms of core dynamics, it is crucial to make the best
use of the new wealth of satellite data that will become available to 
the geomagnetic community, especially with the launch of the SWARM
mission around 2010 \citep{swarmsim}. 

This best use can be achieved in the framework
of data assimilation. In this respect,  geomagnetists are facing challenges similar to
the ones oceanographers were dealing with in the early Nineteen-nineties, with
the advent of operational satellite observation of the oceans. Inasmuch 
as oceanographers benefited from the pioneering work of their atmosphericist colleagues 
(data assimilation is routinely used to improve weather forecasts), geomagnetists
must rely on the developments achieved by the oceanic and atmospheric communities 
to assemble the first bricks of geomagnetic data assimilation. Dynamically 
speaking, the earth's core is closer to the oceans than to the atmosphere.  
The similarity is limited though, since the core is a conducting fluid
whose dynamics are affected by the interaction of the velocity field
with the magnetic field it sustains. These considerations, and their implications
concerning the applicability of sophisticated ocean data assimilation strategies
to the earth's core, will have to be addressed in the future.  
 Today, geomagnetic data assimilation is still in its infancy (see below
for a review of the efforts pursued in the past couple of years). We thus have to 
ask ourselves zero-th order questions, such as: variational or sequential assimilation? 

 In short, one might be naively tempted to say that variational data assimilation (VDA)
 is more versatile than  sequential data assimilation (SDA), at the expense 
of a more involved implementation -for an enlightening introduction to the topic, see \cite{tal97}. 
Through an appropriately defined misfit function, VDA can in principle
answer any question of interest, provided that one resorts
to the appropriate adjoint model. In this paper, 
 we specifically
 address the issue of improving initial conditions to better explain
a data record, and show how this can be achieved,  working with a non-linear, one-dimensional
magneto-hydrodynamic (MHD) model. 
 SDA is more practical, specifically
geared towards better forecasts of the model state, for example 
in numerical weather prediction \citep{tal97}. 
No adjoint model is needed here; the main difficulty lies in  the computational
burden of propagating the error covariance matrix needed to perform the
so-called analysis, the operation by which past information is taken
into account in order to better forecast future model states 
\citep[e.g.][]{brasseur06}.  

Promising efforts in applying SDA concepts and techniques to 
geomagnetism have recently been pursued: \cite{ltk07} have performed
so-called Observing System Simulation Experiments (OSSEs) using 
 a three-dimensional model of the geodynamo, to study in particular
the response (as a function of depth) of the core to surface measurements of 
the normal component of the magnetic field, for different 
 approximations of the above mentioned error covariance matrix.  
 Also, in the context of a simplified one-dimensional MHD model, which retains
part of the ingredients that make the complexity (and the beauty) of the geodynamo, 
\cite{stk07} have applied an optimal interpolation scheme that uses a
Monte-Carlo method to calculate the same matrix, and studied the
response of the system to assimilation for different temporal
and spatial sampling frequencies. Both studies show 
a positive response of the system to SDA (i.e. better forecasts). 

In our opinion, though, SDA is strongly penalized by its formal impossibility to use current 
observations to improve past data records -even if this does not hamper
its potential to produce good estimates of future core states. As said above,
most of the information we have about the core is less that $500$ yr old \citep{gufm}. 
This record contains the signatures of the phenomena responsible for its
short-term dynamics, possibly hydromagnetic waves with periods of several
tens of years \citep{fj2003}. 
Our goal is to explore the VDA route in order 
to see to which extent high-resolution satellite measurements of the
earth's magnetic field can help improve the historical magnetic database,
and identify more precisely physical phenomena responsible for short-term
geomagnetic variations. To tackle this problem, we need a dynamical
model of the high-frequency dynamics of the core, 
and an assimilation strategy. The aim of this paper is to 
reveal the latter, and illustrate it with a simplified 
one-dimensional nonlinear MHD model. 
 Such a toy model, similar to the one used by \cite{stk07}, retains part of the physics, at the
benefit of a negligible computational cost. It enables intensive 
testing of the assimilation algorithm. 

This paper is organized as follows: the methodology we shall pursue in applying variational
data assimilation to the geomagnetic secular variation 
is presented in Sect.~\ref{sec:metho}; its implementation for the 
one-dimensional, nonlinear MHD toy model is described in detail in Sect.~\ref{sec:toy}. 
Various synthetic assimilation experiments are presented in Sect.~\ref{sec:sae}, the results 
of which are summarized and further discussed in Sect.~\ref{sec:dis}. 

\section{Methodology}
\label{sec:metho}
In this section, we outline the bases of variational geomagnetic 
data assimilation, with the mid-term intent of improving the quality 
of the past geomagnetic record using the high-resolution information 
recorded by satellites. We resort to the unified set 
of notations proposed by \cite{icgl97}. What follows is
essentially a transcription of the landmark paper by \cite{tc87} 
 with these conventions, transcription to which we add the 
 possibility
of imposing constraints to the core state itself during the assimilation
process. 

\subsection{Forward model}
Assume we have a prognostic, nonlinear, numerical model $M$ which describes 
the dynamical evolution of the core state at any discrete time
$t_i, i \in \{0, \dots, n\}$. If $\tstep$ denotes the time-step
size, the width of the time window considered here is $t_n-t_0=n \tstep$,
the initial (final) time being $t_0$ ($t_n$). In formal assimilation 
parlance, this is written as
\begin{equation}
\statev_{i+1} = M_i [\statev_{i}],
\label{eq:mod}
\end{equation}
in which $\statev$ is a column vector describing the model state.
 If $M$ relies for instance on the discretization
of the equations governing secular variation with a grid-based approach, 
 this vector contains the values of all the field variables
at every grid point. The secular variation equations could involve terms with a 
known, explicit time dependence, hence the dependence of $M$ on time in Eq.~\eqref{eq:mod}. 
Within this framework, the modeled secular variation is entirely controlled by 
the initial state of the core, $\statev_0$. 
\subsection{Observations}
Assume now that we have knowledge of the true dynamical state of the core
$\statev_i^t$ through databases of observations $\observ$ collected at discrete
locations in space and time: 
\begin{equation}
\observ_i = H_i[\statev_i^t] + \noise_i, 
\end{equation} 
in which $H_i$ and $\noise_i$ are the discrete observation operator and noise, 
 respectively. For GSV, observations consist of (scalar or vector) measurements of the magnetic
field, possibly supplemented by decadal timeseries
of the length of day, since these are related to the angular momentum
of the core \citep{jaultetal1988,blo98}. 
The observation operator is assumed linear and time-dependent: in the context 
 of geomagnetic data assimilation, we can safely anticipate that its dimension 
 will increase dramatically when entering the 
recent satellite era (1999-present). However, $H$ will always produce 
vectors whose dimension is much lower than the dimension of the state itself: this 
fundamental problem of undersampling is at the heart of the development of data assimilation 
strategies. The observational error is time-dependent as well: it is assumed to
have zero mean and we denote its covariance matrix at discrete time $t_i$ by $\obscor_i$. 

\subsection{Quadratic misfit functions}
Variational assimilation aims here at improving the definition of the initial state 
of the core $\statev_0$ to produce modeled observations as close as possible to
the observations of the true state. The distance between observations and predictions
is measured using a quadratic misfit function $J_H$
\begin{equation}
J_H = \sum_{i=0}^{n}  \left[ H_i \statev_{i} - \observ_i \right]^T \ecov_i^{-1}
                      \left[ H_i \statev_{i} - \observ_i \right], 
\label{defjh}
\end{equation}
 in which the superscript `$T$' means transpose.  
In addition to the distance between observations and predictions of the past record, we might
as well wish to try and apply some further constraints on the core state that
we seek, through the addition of an extra cost function $J_C$
\begin{equation}
J_C =  \sum_{i=0}^n \statev_i^T \matconst \statev_i, 
\label{defjc}
\end{equation}
in which $C$ is a matrix which describes the constraint one
would like $\statev$ to be subject to. This constraint can
 originate from some a priori ideas about the physics
 of the true state of the system, and its implication on
the state itself, should this physics not be properly 
accounted for by the model $M$, most likely because of its computational cost. 
In the context of geomagnetic data
assimilation, this a priori constraint can come for
example from the assumption that fluid motions inside the
rapidly rotating core are almost invariant along the direction of earth's
rotation, according to Taylor--Proudman's theorem \cite[e.g.][]{gre90}. 
We shall provide the reader with an example for $C$ when applying these
theoretical concepts to the 1D MHD model (see Sect.~\ref{sec:const}). 

Consequently, we write the total misfit function $J$ as 
\begin{equation}
J = \frac{\alpha_H}{2} J_H + \frac{\alpha_C}{2} J_C,
\label{defj}
\end{equation}
where $\alpha_H$ and $\alpha_C$ are the weights of the 
observational and constraint-based misfits, respectively. 
These two coefficients should be normalized; we will discuss the
normalization in Sect.~\ref{sec:sae}. 

\subsection{Sensitivity to the initial conditions} 
To minimize $J$, we express its 
sensitivity to $\statev_0$, namely 
$\grd_{\statev_0}J$. With our conventions, $\grd_{\statev_0}J$ 
is a row vector, since a change in $\statev_0$, $\delta \statev_0$,
is responsible for a change in $J$, $\delta J$, given by
\begin{equation}
\delta J = \grd_{\statev_{0}} J \cdot \delta \statev_{0}. 
\end{equation}

To compute this gradient, we first introduce the tangent linear
operator which relates a change in 
$\statev_{i+1}$ to a change in the core state at the preceding
discrete time, $\statev_{i}$:
\begin{equation}
\delta \statev_{i+1} = M'_i \delta \statev_{i}.
\end{equation}
The tangent linear operator $M'_i$ is obtained by linearizing 
the model $M_i$ about the state $\statev_i$. 
Successive applications of the above relationship allow us
to relate perturbations of the state vector $\statev_{i} $ at a given model time $t_{i}$
to perturbations of the initial state $\statev_{0}$:
\begin{equation}
\delta \statev_{i} =  \prod_{j=0}^{i-1} M'_j \delta \statev_{0},\forall i \in\{1,\dots,n\}
\label{chaina}
\end{equation}
The sensitivity of $J$ to any $\statev_i$ expresses itself via
\begin{equation}
\delta J = \grd_{\statev_{i}} J \cdot \delta \statev_{i}, 
\label{sensi}
\end{equation}
that is
\begin{equation}
\delta J = \grd_{\statev_{i}} J \cdot \prod_{j=0}^{i-1} M'_j \delta \statev_{0},\ i \in\{1,\dots,n\}. 
\end{equation}
Additionally, after differentiating Eq.~\eqref{defj} using Eqs.~\eqref{defjh} and \eqref{defjc}, we obtain 
$$
\grd_{\statev_{i}} J  = \alpha_H    (H_i \statev_i - \observ_i)^T  \obscor^{-1}_i H_i 
+ \alpha_C \statev_i^T \matconst
 ,\ i \in\{0,\dots,n\}. 
$$
Gathering the observational and constraint contributions to $J$ originating
from every state vector $\statev_i$ finally yields 
\begin{eqnarray*}
\delta J &=& \sum_{i=1}^n \left[ \alpha_H    
         (H_i \statev_i - \observ_i)^T  \obscor^{-1}_i H_i 
             + \alpha_C \statev_i^T \matconst \right] \cdot
            \prod_{j=0}^{i-1} M'_j \delta \statev_{0}  \\
          && + \left[ \alpha_H    (H_0 \statev_0 - \observ_0)^T  \obscor^{-1}_0 H_0 
          + \alpha_C \statev_0^T \matconst \right]  \delta \statev_{0} \\                     
 &=& \left\{ \sum_{i=1}^n \left[ \alpha_H    (H_i \statev_i - \observ_i)^T  \obscor_i^{-1} H_i 
          + \alpha_C \statev_i^T \matconst \right] 
            \prod_{j=0}^{i-1} M'_j \right. \\
        && +  \alpha_H    (H_0 \statev_0 - \observ_0)^T  \obscor_0^{-1} H_0 
           + \alpha_C \statev_0^T \matconst   \Bigg\}   \delta \statev_{0},                     
\end{eqnarray*}
which implies in turn that
\begin{eqnarray}
\grd_{\statev_0} J &=& \sum_{i=1}^n 
            \left[ \alpha_H    (H_i \statev_i - \observ_i)^T  \obscor_i^{-1} H_i + \alpha_C \statev_i^T \matconst \right]
        \prod_{j=0}^{i-1} M'_j \nonumber\\ 
       && + \alpha_H    (H_0 \statev_0 - \observ_0)^T  \obscor_0^{-1} H_0 + \alpha_C \statev_0^T \matconst. 
\label{gradrow}
\end{eqnarray}
\subsection{The adjoint model}
The computation of $\grd_{\statev_0} J $ via Eq.~\eqref{gradrow} is injected in an iterative
method to adjust the initial state of the system to try and minimize $J$. The $l+1$-th step
of this algorithm is given in general terms by
\begin{equation}
\statev_0^{l+1} = \statev_0^{l} - \rho^l \descent^l, 
\end{equation}
in which $\descent$ is a descent direction, and $\rho^l$ an appropriate chosen scalar.
In the case of the steepest descent algorithm, $\descent^l = (\grd_{\statev_0^l} J)^T $, and $\rho^l$
is an a priori set constant. The descent direction is a column vector, hence the need to take
the transpose of $\grd_{\statev_0^l}J$. In practice, 
 the transpose of Eq.~\eqref{gradrow} yields, at the $l$-th step of the algorithm, 
\begin{eqnarray}
\left[\grd_{\statev_0^l} J\right]^T &=&  \sum_{i=1}^n  M'^T_{0} \cdots M'^T_{i-1} 
      \left[\alpha_H H_i^T \obscor^{-1}_i(H_i \statev_i^l - \observ_i) 
     +\alpha_C \matconst \statev_i^l \right] \nonumber \\
     && +\alpha_H  H_0^T  \obscor^{-1}_0 (H_0 \statev_0^l - \observ_0) + \alpha_C \matconst \statev_0^l. 
\end{eqnarray}
Introducing the adjoint variable $\adjoint$, the calculation of $(\grd_{\statev_0^l} J)^T$
is therefore performed practically by integrating the so-called adjoint model
\begin{equation}
\adjoint_{i-1}^l = {M'}^T_{i-1} \adjoint_i^l + \alpha_H H^T_{i-1} \obscor^{-1}_{i-1}(H_{i-1} \statev^l_{i-1} 
               - \observ_{i-1}) + \alpha_C \matconst \statev^l_{i-1}, 
\label{adjm}
\end{equation}
starting from $\adjoint^l_{n+1}=\mitbf{0}$, and going backwards in order to finally estimate 
\begin{equation}
(\grd_{\statev_0^l} J)^T = \adjoint^l_0.  
\end{equation}
Equation \eqref{adjm} is at the heart of variational data assimilation \citep{tal97}. Some
 remarks and comments concerning this so-called adjoint equation are in order: 
\begin{enumerate}
\item It requires to implement the transpose of the tangent linear operator, the so-called 
      adjoint operator, ${M'}^T_{i}$. If the discretized forward model is 
      cast in terms of matrix-matrix and/or matrix-vector products, then this implementation
      can be rather straightforward (see Sect.~\ref{sec:toy}). Still, for realistic
      applications, deriving the discrete adjoint equation can be rather convoluted 
      \cite[e.g.][Chap. 4]{ben02}. 
\item The discrete adjoint equation (Eq.~\ref{adjm}) is based on the already discretized model of the
      secular variation. Such an approach is essentially motivated by practical reasons, assuming that
      we already have a numerical model of the geomagnetic secular variation at hand. We should mention
      here that a similar effort can be performed at the continuous level, before discretization. 
      The misfit can be defined at this level; the calculus of its variations gives then rise
      to the  Euler--Lagrange equations, one of which being the continuous backward, or adjoint,  
      equation. One could then simply discretize this equation, using the same numerical
      approach as the one used for the forward model, and use this tool to adjust $\statev_0$.
      According to \cite{ben02}, though, the ``discrete adjoint equation" is not the
      ``adjoint discrete equation", the former breaking adjoint symmetry, which results in
      a solution being suboptimal \citep[][\S~4.1.6]{ben02}. 
\item Aside from the initial state $\statev_0$, one can in principle add model parameters 
      ($\mparam$, say) as adjustable variables, and invert jointly for $\statev_0$ and $\mparam$, at the expense
      of expressing the discrete sensitivity of $J$ to $\mparam$ as well. For geomagnetic VDA,
      this versatility might be of interest, in order for instance to assess the importance 
      of magnetic diffusion over the time window of the historical geomagnetic record. 
\item The whole sequence of core states $\statev_{i}^l, i \in \{0,\dots,n \}$, has to be kept in memory. 
      This memory requirement can become quite significant when considering dynamical 
      models of the GSV. Besides, even if the computational cost of the adjoint model is
      by construction equivalent to the cost of the forward model, the variational 
      assimilation algorithm presented here is at least one or two orders of magnitude
      more expensive than a single forward realization, because of the number of iterations  
      needed to obtain a significant reduction of the misfit function. 
      When tackling `real' problems in the future (as opposed to the illustrative
      problem of the next sections), memory and CPU time constraints might make it
      necessary to lower the resolution of the forward (and adjoint) models, by
      taking parameters values further away from the real core. A constraint 
      such as the one imposed through Eq.~\eqref{defjc} can then appear as a way to ease the pain and not to 
      sacrifice too much physics, at negligible extra computational cost. 
\end{enumerate} 
We give a practical illustration of these ideas and concepts in the next two sections. 

\section{Application to a one-dimensional nonlinear MHD model}
\label{sec:toy}
We consider a conducting fluid, whose state is fully characterized
by two scalar fields, $u$ and $b$. Formally, 
$b$ represents the magnetic field (it can be observed), and
$u$ is the velocity field (it is invisible). 
\subsection{The forward model}

\subsubsection{Governing equations}
The conducting fluid has density $\rho$,
 kinematic viscosity $\nu$, electrical conductivity $\sigma$, 
  magnetic diffusivity $\eta$, and magnetic permeability
 $\mu$ ($\eta=1/\mu \sigma$). Its pseudo-velocity $u$ 
and pseudo-magnetic field $b$ are both scalar fields, defined
over a domain of length $2L$, $[-L,L]$. We refer to pseudo fields here
since these fields are not divergence-free. If they were so, they would have
to be constant over the domain, which would considerably limit their
interest from the assimilation standpoint. Bearing this remark in
mind, we shall omit the `pseudo' adjective in the remainder of this study. 

We choose $L$ as the length scale, the magnetic diffusion time scale
$L^2/\eta$ as the time scale,  
$B_0$ as the magnetic field scale, and $B_0/\sqrt{\rho \mu}$ as the 
velocity scale (i.e. the Alfv\'en wave speed).  
Accordingly, the evolution of $u$ and $b$ is controlled by the following set
of non-dimensional equations: 
\begin{eqnarray}
\forall (x,t) \in ]-1,1[ \times [0,T], \nonumber \\
\partial_t u + S \ u \partial_x u &=&  S\  b \partial_x b + Pm\partial_x^2 u \label{mom}, \\
\partial_t b +S \ u \partial_x b &=&  S \ b \partial_x u +  \partial_x^2 b, \label{ind}
\end{eqnarray}
supplemented by the boundary and initial conditions
\begin{eqnarray}
u(x,t) &=& 0 \mbox{ if } x = \pm 1,  \\
b(x,t) &=& \pm 1 \mbox{ if } x = \pm 1,  \\
&+& \mbox{ given } u(\cdot,t=0), b(\cdot,t=0).  
\end{eqnarray}
Eq.~\eqref{mom} is the momentum equation: the rate of change of the velocity is controlled
by advection, magnetic forces and diffusion. Similarly, in the
induction equation \eqref{ind}, the rate of change of the magnetic field results from
the competition between inductive effects and ohmic diffusion. 

Two non-dimensional numbers define this system,
$$
S = \sqrt{\mu/\rho} \sigma B_0 L, 
$$
which is the Lundquist number (ratio of the magnetic diffusion time scale
  to the Alfv\'en time scale), and 
$$
 Pm = \nu/\eta,
$$
which is the magnetic Prandtl number, a material property very small for liquid metals - $Pm~\sim~10^{-5}$ 
for earth's core \citep[e.g.][]{jpp88}. 

\subsubsection{Numerical model}
\label{sec:num}
\begin{figure}
\centerline{\includegraphics[width=.5\linewidth]{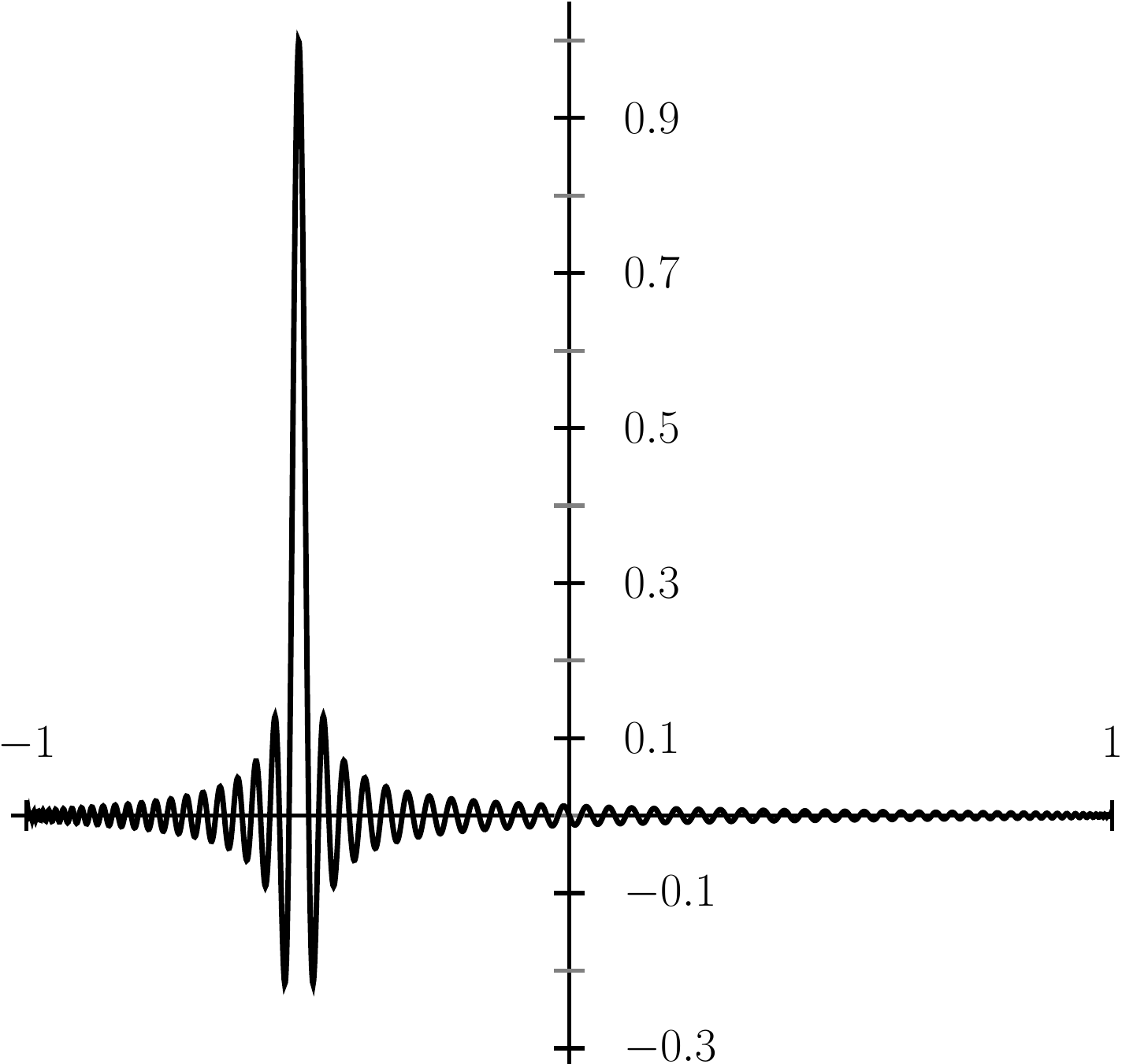}}
\caption{\label{fig:bf} An example of a basis function used to discretize the MHD 
         model in space. This particular Lagrangian interpolant, $
         h_{50}^{150}$, is obtained for a polynomial order $N=150$, and 
it is attached to the 51st Gauss--Lobatto--Legendre point.}
\end{figure}

Fields are discretized in space using one Legendre spectral element of order $N$. 
In such a framework, basis functions are the Lagrangian interpolants $h_i^N$defined
over the collection of $N+1$ Gauss--Lobatto--Legendre (GLL) points $\xi_i^N, 
i \in \{0,\dots,N\}$ \citep[for a comprehensive description
of the spectral element method, see][]{dfm02}. Figure~\ref{fig:bf} shows
 such a basis function for $i=50,N=150$. Having basis functions defined 
everywhere over $[-1,1]$ makes it straightforward to define numerically the observation
operator $H$ (see Sect. \ref{sec:obstrue}). We now drop the superscript $N$ for the sake of 
 brevity. The semi-discretized velocity and magnetic fields are column vectors,
denoted with bold fonts 
\begin{eqnarray}
\udisc(t)&=&\left[u(\xi_0=-1,t),u(\xi_1,t),\dots,u(\xi_N=1,t) \right]^T, \\
\bdisc(t)&=&\left[b(\xi_0=-1,t),b(\xi_1,t),\dots,b(\xi_N=1,t) \right]^T.
\end{eqnarray}
Discretization is performed in time with a semi-implicit finite-differencing
scheme of order $1$, explicit for nonlinear terms, and implicit for 
diffusive terms. As in the previous section, assuming that $\Delta t$
is the time step size, we define $t_i = i \tstep, \udisc_i = \udisc(t=t_i), 
                                  \bdisc_i = \bdisc(t=t_i), i \in \{0,\dots,n\}.$ 
As a result of discretization in both space and time, the model is
advanced in time by solving the following algebraic system 
\begin{equation}
\ub{i+1} = \left[
           \begin{array}{cc}
            \helmu^{-1}   &  0   \\
            0  &  \helmb^{-1}    \\
           \end{array}
           \right] 
           \fub{i},
\end{equation}
where \begin{eqnarray}
        \helmu &=& \massm/\tstep +  Pm \stiff,  \\
        \helmb &=& \massm/\tstep +     \stiff, \\
        \fudisc_{u,i} &=&  \massm \left(\udisc_i/\tstep 
                                       -S \udisc_i \had \deriv \udisc_i  
                                       +S \bdisc_i \had \deriv \bdisc_i  \right ),  \\ 
        \fbdisc_{b,i} &=&  \massm \left(\bdisc_i /\tstep   
                                        -S \udisc_i \had \deriv \bdisc_i
                                        +S \bdisc_i \had \deriv \udisc_i \right),  
      \end{eqnarray}
are the Helmholtz operators acting on velocity field and the magnetic 
field, and the forcing terms for each of these two, respectively. 
We have introduced the following definitions :
\begin{itemize}
  \item $\massm$, which  is the diagonal mass matrix, 
  \item $\stiff$, which is the so-called  stiffness matrix (it is symmetric definite positive),
  \item $\had$, which denotes the Hadamard product: $(\bdisc\had \udisc)_k = (\udisc\had \bdisc)_k = b_k u_k$, 
  \item and $\deriv$, the so-called derivative matrix 
        \begin{equation}\deriv_{ij} = \frac{dh^{N}_{i}}{dx}|_{x=\xi_j}, \end{equation}
\end{itemize}
the knowledge of which is required to evaluate the nonlinear terms. Advancing
in time requires to invert both Helmholtz operators, which we do directly resorting
to standard linear algebra routines \citep{laug}. 
Let us also bear in mind that the Helmholtz operators
are symmetric (i.e. self-adjoint).

In assimilation parlance, and according to the conventions introduced in the previous section, 
the state vector $\statev$ is consequently equal to $[\udisc,\bdisc]^T$, and its dimension is
$s=2(N-1)$ (since the value of both the velocity and magnetic fields are prescribed
on the boundaries of the domain). 

\subsection{The true state}
\label{true}
\begin{figure*}
\centerline{\includegraphics[width=.8\linewidth]{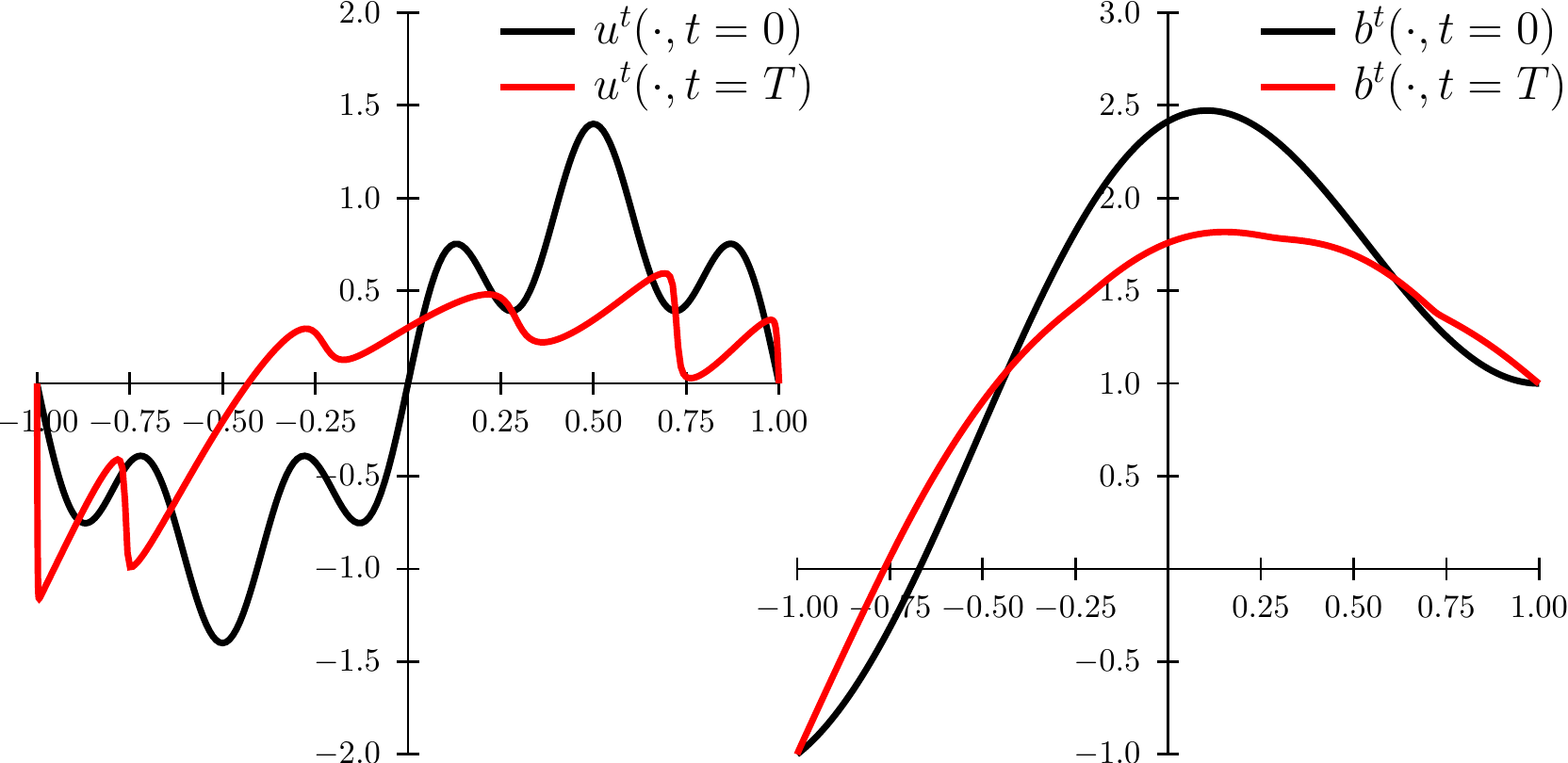}}
\caption{\label{fig:true} The true state used for synthetic variational assimilation experiments. Left: the 
first, $t=0$ (black) and last, $t=T$ (red) velocity fields. Right: the first, $t=0$ (black) and last, 
  $t=T$  (red) magnetic fields.} 
\end{figure*}

Since we are dealing in this paper with synthetic observations, it is necessary
to define the true state of the 1D system as the state obtained via the integration
of the numerical model defined in the preceding paragraph, for a given set of 
initial conditions, and specific values of the Lundquist and magnetic Prandtl
numbers, $S$ and $Pm$. The true state (denoted with
the superscript `$t$') will always refer
to the following initial conditions 
\begin{eqnarray}
u^t(x,t=0) &=&  \sin(\pi x) + (2/5) \sin (5 \pi x), \label{ut}\\
b^t(x,t=0) &=&  \cos(\pi x) + 2 \sin [\pi(x+1)/4], \label{bt}
\end{eqnarray}
along with  $S=1$ and $Pm=10^{-3}$. The model is integrated forward in time
until $T=0.2$ (a fifth of a magnetic diffusion time). The polynomial order used
to compute the true state is $N=300$, and the time step size $\Delta t=2\ 10^{-3}$. 
Figure \ref{fig:true} shows the velocity (left) and magnetic field (right)
at initial (black curves) and final (red curves) model times. 
The low value of the magnetic Prandtl number $Pm$ reflects itself in the sharp velocity boundary layers
that develop near the domain boundaries, while the magnetic field exhibits in contrast a 
smooth profile (the magnetic diffusivity being three orders of magnitude larger than
the kinematic viscosity). To properly resolve these  Hartmann boundary layers there must be
enough points in the vicinity of the domain boundaries: we benefit here from the clustering of GLL 
points near the boundaries \citep{dfm02}. 
Besides, even if the magnetic profile is very smooth, one can nevertheless point out here and
there kinks in the final profile. These kinks are associated with sharp velocity gradients 
(such as the one around $x=0.75$) and are a consequence of the nonlinear $b \partial_x u$ term
in the induction Eq.~\eqref{ind}. 
\subsection{Observation of the true state}
\label{sec:obstrue}
In order to mimic the situation relevant for the earth's core and geomagnetic secular variation, 
assume that  we have knowledge of $b$ at discrete locations in space and time, and that the 
 velocity $u$ is not measurable. 
For the sake of generality, observations of $b$ are not necessarily made
at collocation points, hence the need to define a spatial observation operator $\hspa_i$ (at discrete
 time $t_i$) consistent
with the numerical approximation introduced above. If $\nsta_i$ denotes the number
of virtual magnetic stations at time $t_i$, and $\xi^o_{i,j}$ their locations ($j \in \{1,\dots, \nsta_i \}$),
$\hspa_i$ is a rectangular $\nsta_i \times (N+1) $ matrix, whose coefficients write
\begin{equation}
\hspa_{i,jl} = h_l^N(\xi_{i,j}^o). 
\end{equation}
A database of magnetic observations $\observ_i=\bdisc^o_i$ is therefore produced at discrete time $t_i$ via
the matrix-vector product \begin{equation}
                          \bdisc^o_i = \hspa_i \bdisc_i^t.  
                          \end{equation}
Integration of the adjoint model also requires the knowledge of the transpose of the observation operator
(Eq.~\ref{adjm}), the construction of which is straightforward according to the previous definition.  
To construct the set of synthetic observations, we take for simplicity the observational noise to be zero. 
During the assimilation process, we shall assume that estimation
errors are uncorrelated,  and that the level of confidence is the same for each virtual observatory. 
 Consequently,   
\begin{equation}
\obscor_i=\massio,
\end{equation}
in which $\massio$ is the $\nsta_i \times \nsta_i$ identity matrix, throughout the numerical experiments. 

 As an aside, let us notice that magnetic observations could equivalently consist of an (arbitrarily truncated) 
 set of spectral coefficients, resulting from the expansion of the magnetic field on the basis of 
 Legendre polynomials. Our use of stations is essentially motivated by the fact that our forward
 model is built in physical space. 
 For real applications, a spectral approach is interesting since it can naturally 
 account for the potential
 character of the field in a source-free region; however, it is less amenable 
 to the spatial description of observation errors, if these do not vary smoothly.

\subsection{The adjoint model}
\subsubsection{The tangent linear operator}
As stated in the the previous section, the tangent linear operator $M'_i$ at discrete time $t_i$
is obtained at the discrete level by linearizing the model about the current solution $(\udisc_i,\bdisc_i)$. 
By perturbing these two fields
\begin{eqnarray}
\udisc_i \rightarrow \udisc_i + \delta \udisc_i,\\
\bdisc_i \rightarrow \bdisc_i + \delta \bdisc_i, 
\end{eqnarray}
we get (after some algebra)  
$$
\dub{i+1} = \left[\begin{array}{cc}
                  \mata_i &   \matb_i \\
                  \matc_i &   \mate_i 
                 \end{array} \right]\dub{i}
$$
having introduced the $(N+1)^2$ following matrices 
\begin{eqnarray}
 \mata_i &=& \helmu^{-1} \massm \left( \massi/\tstep -S \deriv \udisc_i \had - S \udisc_i \had \deriv \right), \\
 \matb_i &=& \helmu^{-1} \massm \left(  S \bdisc_i \had \deriv + S \deriv \bdisc_i \had \right),\\
 \matc_i &=& \helmb^{-1} \massm \left( -S \deriv \bdisc_i \had -S \bdisc_i \had \deriv \right),\\
 \mate_i &=& \helmb^{-1} \massm \left( \massi/\tstep -S  \udisc_i \had \deriv + S \deriv \udisc_i \had \right).  
\end{eqnarray}
Aside from the $(N+1)^2$ identity matrix $\massi$,  matrices and notations appearing in these definitions have already been introduced in \S \ref{sec:num}. 
In connection with the general definition introduced in the previous section, 
$\delta \statev_{i+1} = M'_i \delta \statev_{i}$,
 $M'_i$ is the block matrix 
\begin{equation}
M'_i = \left[\begin{array}{cc}
                  \mata_i &   \matb_i \\
                  \matc_i &   \mate_i 
                 \end{array}
       \right].
\label{eq:tlo}
\end{equation}
\subsubsection{Implementation of the adjoint equation}
The sensitivity of the model to its initial
conditions is computed by applying the adjoint operator, $M_i'^T$, to 
the adjoint variables - see Eq.~\eqref{adjm}. According to Eq.~\eqref{eq:tlo}, one gets
\begin{equation}
M_i'^T = \left[\begin{array}{cc}
                  \mata_i^T &   \matc_i^T \\
                  \matb_i^T &   \mate_i^T 
                 \end{array}
       \right],
\label{eq:adjb}
\end{equation}
with each transpose given by 
\begin{eqnarray}
 \mata_i^T &=& \left( \massi/\tstep -S \udisc_i \had \deriv^T - S \deriv^T \udisc_i \had  \right) 
               \massm \helmu^{-1}, \\
 \matb_i^T &=& \left(  S \deriv^T \bdisc_i \had  + S \bdisc_i \had \deriv^T  \right)\massm \helmu^{-1}, \\
 \matc_i^T &=& \left(  - S \bdisc_i \had \deriv^T -S \deriv^T \bdisc_i \had \right)\massm \helmb^{-1}, \\
 \mate_i^T &=& \left(\massi/\tstep   -S \deriv^T \udisc_i \had  + S \udisc_i \had \deriv^T \right)\massm \helmb^{-1}. 
\end{eqnarray}
In writing the equation in this form, we have used the symmetry properties 
of the Helmholtz and mass matrices, and introduced the
transpose of the derivative matrix, $D^T$. Programming the adjoint model is very similar to programming
the forward model, provided that one has cast the latter in terms of matrix-matrix, matrix-vector, and 
Hadamard products. 

\section{Synthetic assimilation experiments}
\label{sec:sae}
\begin{figure*}
\centerline{\includegraphics[width=.8\linewidth]{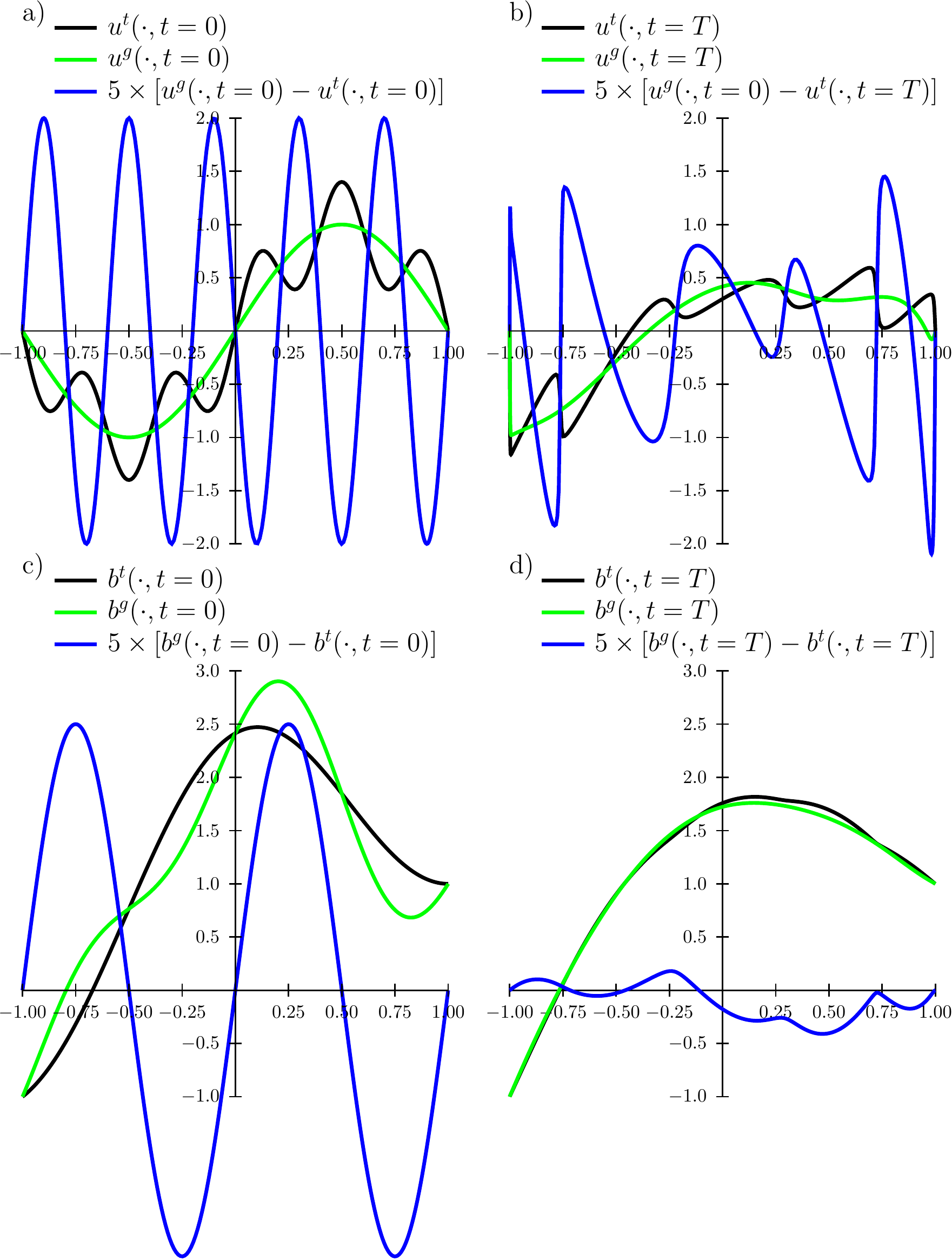}}
\caption{\label{fig:gvst} Initial guesses used for the variational assimilation experiments,
 plotted against the corresponding true state variables. Also plotted is
five times the difference between the two. a: velocity at time $0$. b:  
velocity at final time $T$. c: magnetic field at time $0$. d: magnetic field
  at final time $T$. 
In each panel, the true state is plotted with the black line, the guess
with the green line, and the magnified difference with the blue line. 
  }
\end{figure*}

Having all the numerical tools at hand, we start out by assuming that we have imperfect knowledge 
of the initial model state, through an initial guess $\statev_0^g$, with the model parameters
and resolution equal to the ones that helped us define the true state of \S \ref{true}. 
We wish here to quantify how assimilation of observations can help improve the knowledge 
of the initial (and subsequent) states, with particular emphasis 
on the influence of spatial and temporal sampling. In the series of results reported 
in this paragraph, the initial guess at model initial time is :
\begin{eqnarray}
u^g(x,t=0) &=&  \sin(\pi x), \label{ug}\\
b^g(x,t=0) &=&  \cos(\pi x) + 2 \sin [\pi(x+1)/4] + (1/2) \sin (2 \pi x). \label{bg}
\end{eqnarray}
With respect to the true state at the initial time, the first guess is missing the
 small-scale component of $u$, i.e. the second term on the right-hand side of  
 Eq.~\eqref{ut}. In addition, our estimate of $b$ has an extra parasitic large-scale 
 component (the third term on the right-hand side of Eq.~\eqref{bg}), 
a situation that could occur when dealing with the GSV, for which the importance
of unmodeled small-scale features has been recently put forward given
the accuracy of satellite data \citep{eh05}. 
Figure \ref{fig:gvst} shows the initial and final $u^g$ and $b^g$, along with
$u^t$ and $b^t$ at the same epochs for comparison, and the difference between
the two, multiplied by a factor of five.
Differences in $b$ are not pronounced. Over the time window considered here,
the parasitic small-scale component has undergone considerable diffusion.  
To quantify the differences between the true state and the guess, we resort 
to the $L_2$ norm  
$$\ltwo{f}= \sqrt{\int_{-1}^{+1} f^2 dx}, $$ 
and define the relative magnetic and fluid errors at time $t_i$ by
\begin{eqnarray}
\eb_i &=& \ltwo{b^t_i-b^f_i}/\ltwo{b^t_i},\\
\eu_i &=& \ltwo{u^t_i-u^f_i}/\ltwo{u^t_i}.  
\end{eqnarray}
The initial guess given by Eqs.~\eqref{ug}\eqref{bg} is characterized 
by the following errors: 
$\eb_0 = 21.6 \%, \eb_n=2.9 \%, \eu_0=37.1 \%,  $ and $\eu_n=37.1 \%$. 

\subsection{Improvement of the initial guess with no a priori constraint on the state}
\subsubsection{Regular space and time sampling}
Observations of $b^t$ are performed at $\nsta$ virtual observatories which are equidistant
in space, at a number of epochs $nt$ evenly distributed over the time interval. 
Assuming no a priori constraint on the state, we set
$\alpha_C = 0$ in Def. \ref{defj}. The other constant $\alpha_H = 1/(nt\nsta).$

The minimization problem is tackled by means
of a conjugate gradient algorithm, \`a la Polak--Ribi\`ere \citep{she94}.
Iterations are stopped either when the initial misfit has decreased by 8 orders
of magnitude, or when the iteration count exceeds 5,000. In most cases, the 
latter situation has appeared in our simulations. A typical minimization is
characterized by a fast decrease in the misfit during the first few tens of
iterations, followed by a slowly decreasing (almost flat) behaviour. Even if
the solution keeps on getting better (i.e. closer to the synthetic reality)
during this slow convergence period, practical considerations (having in mind
the cost of future geomagnetic assimilations) prompted us to stop the minimization. 

\begin{figure*}
\centerline{\includegraphics[width=.8\linewidth]{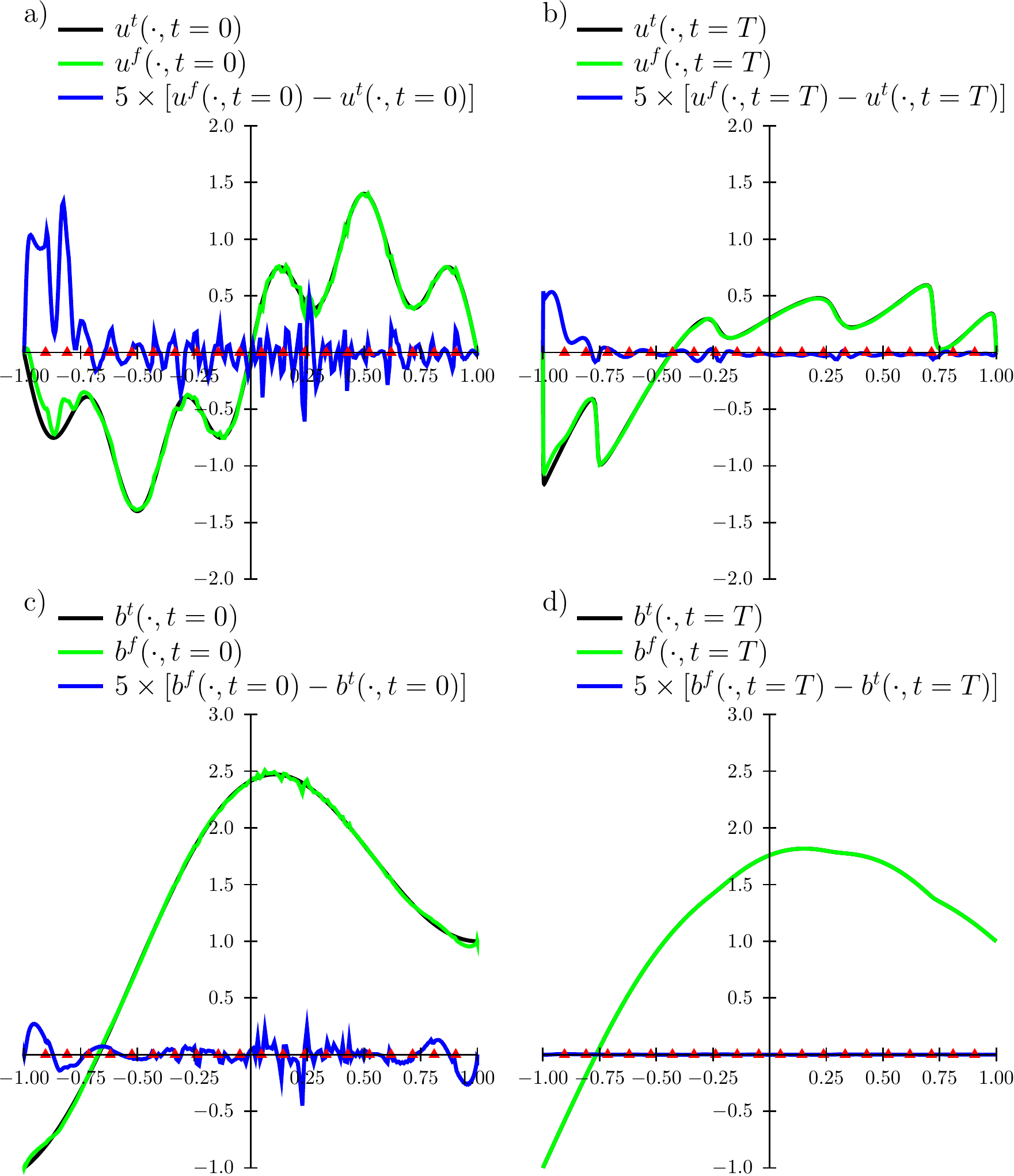}}
\caption{\label{fig:typ} Synthetic assimilation results. a):  
         velocity at initial model time $t=0$. b): velocity at final time $t=T$. 
         c): magnetic field at initial time $t=0$. d): magnetic field at final time $t=T$.
         In each panel, the true field is plotted in black, the assimilated field (starting
         from the guess shown in Fig. \ref{fig:gvst}) in green, and the difference between 
         the two, multiplied by a factor of 5, is shown in blue.  
         The red triangles
         indicate the location of the $\nsta$ virtual magnetic observatories ($\nsta=20$ in 
         this particular case).}
\end{figure*}
A typical example of a variational assimilation result is shown in 
Fig. \ref{fig:typ}. In this case, $\nsta=20$ and $nt=20$. 
The recovery of the final magnetic field $b_n$ is
excellent (see Fig.  \ref{fig:typ}d), the 
relative $L_2$ error being $1.8\ 10^{-4}$. The benefit here
is double: first, the network of observatories is dense 
enough to sample properly the field,  and second, a measurement is made exactly
at this discrete time instant, leaving no time for error fields
to develop. When the latter is possible, the recovered fields can
be contaminated by small-scale features, that is features that have
length scales smaller than the spatial sampling scale. We see this happening
in Fig. \ref{fig:typ}c), in which the magnified difference between the recovered 
and true $b_0$, shown in blue, appears indeed quite spiky; $\eb_0$
has still decreased from an initial value of $21.6 \%$ (Fig. \ref{fig:gvst}c) down to $1.2 \%$. 
Results for velocity are shown in Figs. \ref{fig:typ}a and \ref{fig:typ}b. The recovered
velocity is closer to the true state than the initial guess: this is the expected benefit 
from the nonlinear coupling between velocity and magnetic field in Eqs.~\eqref{mom}-\eqref{ind}. 
 The indirect knowledge we have of $u$, through the observation of $b$,
is sufficient to get better estimates of this field variable.  
At the end of the assimilation process, $\eu_0$ and $\eu_n$, which were 
approximately equal to $37 \%$ with the initial guess, have been brought down to 
$8.2$ and $4.7$ \%, respectively. The velocity at present time (Fig. $\ref{fig:typ} $)
is remarkably close to the true velocity, save for the left boundary layer sharp structure,
which is undersampled (see the distribution of red triangles).  
\begin{figure*}
\centerline{\includegraphics[width=\linewidth]{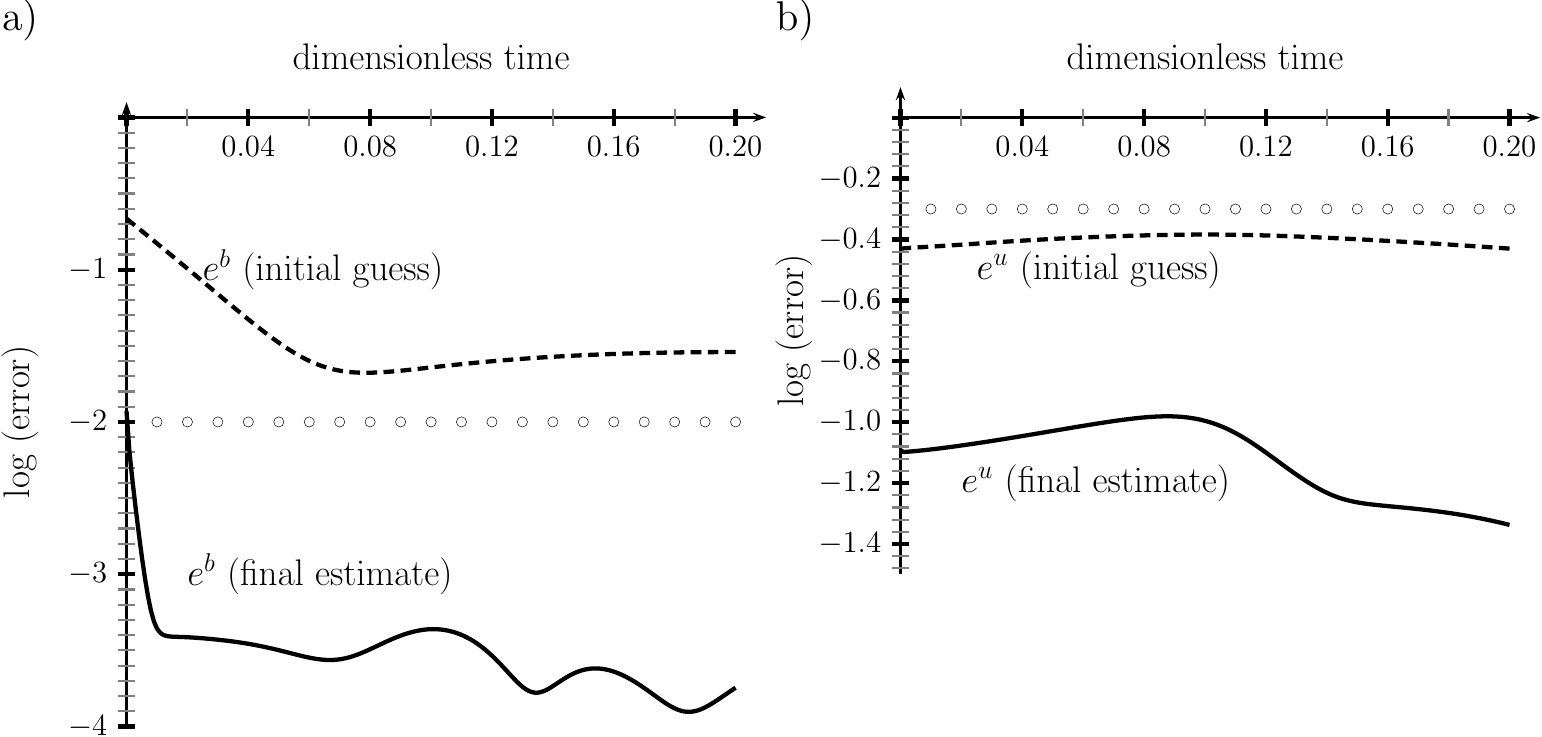}}
\caption{\label{fig:dyne} Dynamical evolution of $L_2$ errors (logarithmic value)
         for the magnetic field (a) and the fluid velocity (b). 
         Dashed lines : errors for initial guesses.  Solid lines : errors after 
         variational assimilation. Circles represent instants are which magnetic
         observations are made. In this particular case, $nt=20$ and $\nsta=20$.}
\end{figure*}
We further document the dynamical evolution of $L_2$ errors by plotting on Fig.~\ref{fig:dyne}
the temporal evolution of $\eb$ and $\eu$ for this particular configuration. Instants at which
observations are made are represented by circles, and the temporal evolution of the guess
errors are also shown for comparison. The guess for the initial magnetic field is characterized
by a decrease of the error that occurs over $\approx .1$ diffusion time scale, that is 
roughly the time it takes for most of the parasitic small-scale error component to diffuse
away, the error being then dominated at later epochs by advection errors, originating from
errors in the velocity field. The recovered magnetic field (Fig.~\ref{fig:dyne}a, solid line),
is in very good agreement with the true field as soon as measurements are available ($t\ge 1\%$
of a magnetic diffusion time, see the circles on Fig.~\ref{fig:dyne}a). Even though no measurements
are available for the initial epoch, the initial field has also been corrected significantly,
as discussed above. In the latter parts of the record, oscillations in the magnetic error field are
present -they disappear if the minimization is pushed further (not shown). 

The unobserved velocity field does not exhibit such a drastic reduction in error as soon
as observations are available (Fig.~\ref{fig:dyne}b, solid line). Still, it is worth noticing
that the velocity error is significantly smaller in the second part of the record, in 
connection with the physical observation that most of the parasitic small-scale component
of the field has decayed away (see above): advection errors dominate in determining the
time derivative of $b$ in Eq.~\eqref{ind}, leaving room for a better assessment of the value
of $u$.  For other cases (different $\nsta$ and $nt$), we find a similar behaviour (not
shown). We comment on the effects of an irregular time sampling on the above observations in 
section \ref{irregtime}. 

\begin{figure*}
\centerline{\includegraphics[width=.8\linewidth]{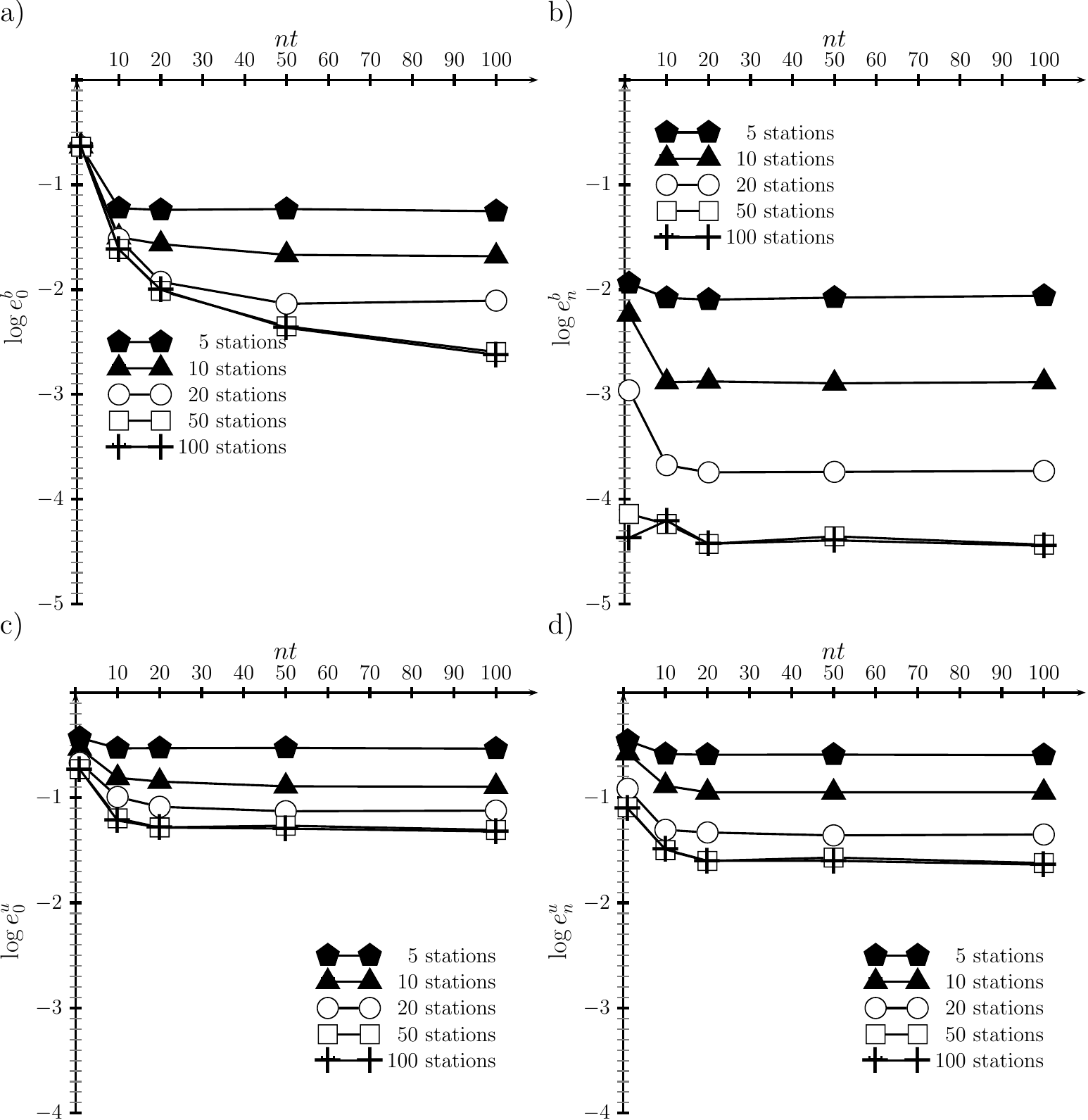}}
\caption{\label{fig:sys} Systematic study of the response of the one-dimensional 
         MHD system to variational assimilation. Shown are the logarithms
         of $L_2$ errors for the $t=0$ (a) and $t=T$ (b) magnetic field,
         and the $t=0$ (c) and $t=T$ (d) velocity field, versus the 
         number of times observations are made over [0,T], $nt$, using
         spatial networks of varying density ($\nsta=5,10,20,50$ and $100$).}
\end{figure*}

Having in mind what one gets in this particular $(nt,\nsta)$ configuration, 
we now summarize in Fig.~\ref{fig:sys} results obtained by varying
systematically these $2$ parameters. After assimilation,
the logarithmic value of the $L_2$ velocity and magnetic field errors, at the initial and final stages
 ($i=0$ and $i=n$),
are plotted versus $nt$, using $\nsta=5, 10, 20, 50,$ and $100$ virtual 
magnetic stations. As far as temporal sampling is concerned, $nt$
can be equal to $1$ (having one observation at present time only), $10$,
$20$, $50$ or $100$. 
Inspection of Fig.~\ref{fig:sys} leads us to make the following comments: 
\begin{itemize}
\item Regarding $b$ :
\begin{itemize}
\item 50 stations are enough to properly sample the magnetic field in space. 
      In this case $nt=1$ is sufficient to properly determine $\bdisc_n$,
      and no real improvement is made when increasing $nt$ (Fig.~\ref{fig:sys}b). 
      During the iterative process, the value of the field is brought to its observed 
      value at every station of the dense network, and this is it: no dynamical information
      is needed. 
\item When, on the other hand, spatial sampling is not good enough, information
      on the dynamics of $b$ helps improve partially its knowledge at present time. For instance, 
      we get a factor of $5$ reduction in $\eb_n$ with $\nsta=20$, going from $nt=1$
      to $nt=10$ (Fig.~\ref{fig:sys}b, circles). The improvement then stabilizes
      upon increasing $nt$: spatial error dominates. 
\item This also applies for the initial magnetic field  $\bdisc_0$, see Fig.~\ref{fig:sys}a.
      As a matter of fact, having no dynamical information about $b$ ($nt=1$) precludes any 
      improvement on $\bdisc_0$, for any density of the spatial network. Improvement occurs
      for $nt>1$. If the spatial coverage is good enough ($\nsta>50$), no plateau is reached,
      since the agreement between the assimilated and true fields keeps on getting better, 
      as it should. 
\end{itemize}
\item Regarding $u$ :
\begin{itemize}
\item The recovered $u$ is always sensitive to spatial
      resolution, even for $nt=1$ (Figs.~\ref{fig:sys}c and \ref{fig:sys}d). 
\item If $nt$ is increased, the error decreases and reaches a plateau which is again 
      determined by spatial resolution. This holds for $\eu_0$ and $\eu_n$. For the
      reason stated above, $\udisc_n$ is better known than $\udisc_0$. The error
      is dominated in both cases by a poor description of the left boundary layer
      (see the blue curves in Figs.~\ref{fig:typ}a and \ref{fig:typ}b). The gradient associated with  
      this layer is not sufficiently well constrained by magnetic observations 
      (one reason being that the magnetic diffusivity is three times larger
      than the kinematic viscosity). Consequently, we can speculate that
      the error made in this specific region at the final time is retro-propagated and amplified going
      backwards in time, through the adjoint equation, resulting in $\eu_0>\eu_n$.
\end{itemize}
\end{itemize}

\subsubsection{Irregular spatial sampling}
We have also studied the effect of an irregular spatial sampling by performing a suite
of simulations identical to the ones described above, save that we assumed that
stations were only located in the left half of the domain (i.e. the $[-1,0]$ segment). 

\begin{figure*}
\centerline{\includegraphics[width=.8\linewidth]{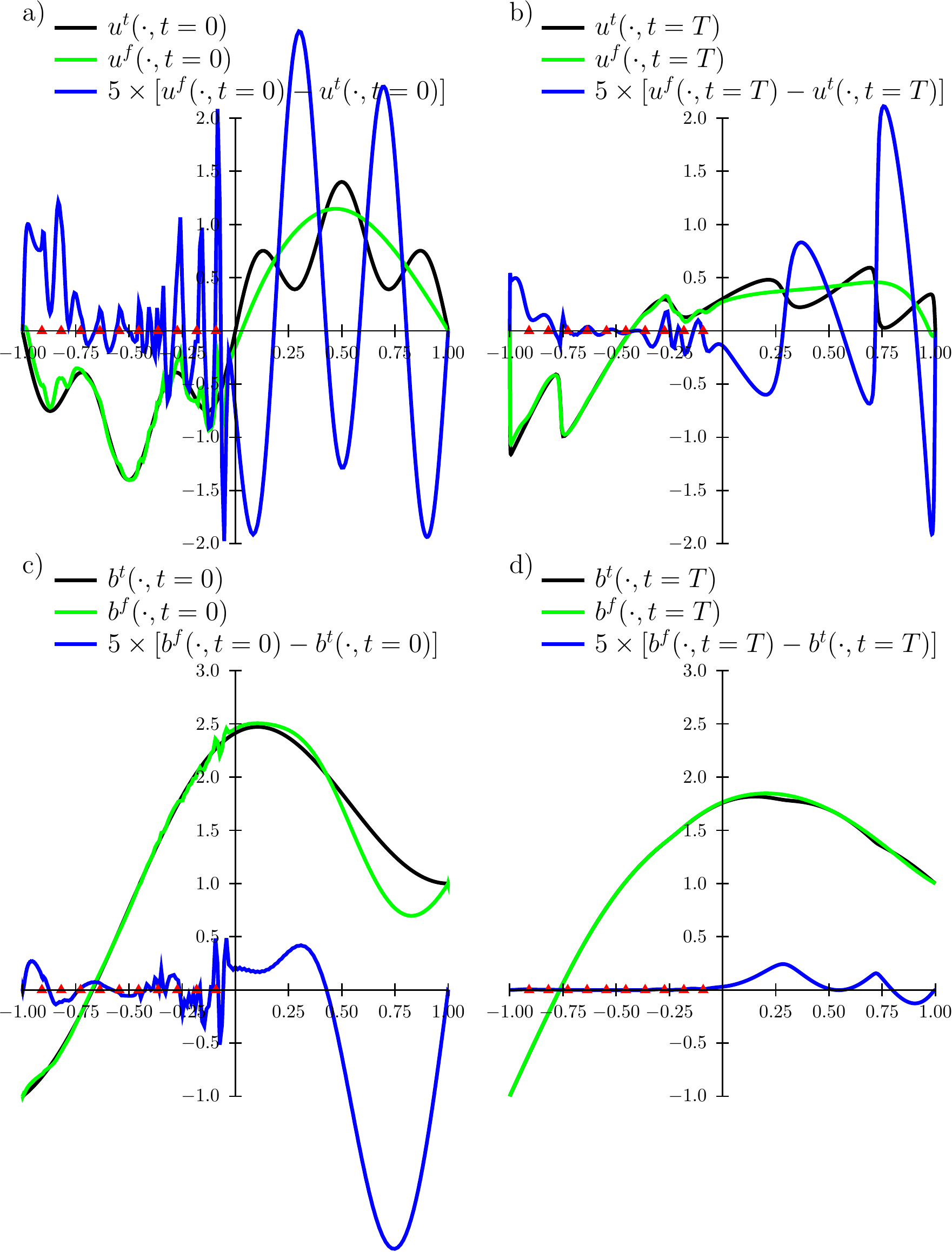}}
\caption{\label{fig:bias} Synthetic assimilation results obtained with an asymmetric network of virtual
         observatories (red triangles). Other model and assimilation parameters 
         as in Fig.~\ref{fig:typ}. a):  
         velocity at initial model time $t=0$. b): velocity at final model time $t=T$. 
         c): magnetic field at $t=0$. d): magnetic field at $t=T$. 
         In each panel, the true field is plotted in black, the assimilated field 
          in green, and the difference between
         the two, multiplied by a factor of $5$, is shown in blue.}
\end{figure*}

The global conclusion is then the following: assimilation results in an improvement of estimates 
of $b$ and $u$ in the sampled region, whereas no benefit is visible in the unconstrained region. 
To illustrate this tendency (and keep a long story short), we
only report in Fig.~\ref{fig:bias} the recovered $u$ and $b$ for $(\nsta,nt)= (10,20)$, which corresponds
to the ``regular" case depicted in Fig. \ref{fig:typ}, deprived from its $10$ stations
located in $[0,1]$. 
The lack of transmission of information from the left-hand side of the domain to its
right-hand side is related to the short duration of model integration ($0.2$ magnetic diffusion time, which
corresponds to $0.2$ advective diffusion time with our choice of $S=1$).  We shall comment 
further on the relevance of this remark for the assimilation of the historical geomagnetic
secular variation in the discussion. 

The lack of observational constraint on the right-hand side  of the domain results sometimes
in final errors larger than the initial ones (compare in particular Figs.~\ref{fig:bias}a 
and \ref{fig:bias}b, with Figs.~\ref{fig:typ}a and \ref{fig:typ}b).  

We also note large error oscillations located at the interface between 
the left (sampled) and right (not sampled) regions, particularly at initial model
time (Figs.~\ref{fig:bias}a and \ref{fig:bias}c). The contrast in spatial
 coverage is likely to be the cause of these oscillations (for which we do not
have a formal explanation); this type of behaviour should be kept in mind for future
geomagnetic applications. 
\subsubsection{Irregular time sampling}
\label{irregtime}
\begin{figure*}
\centerline{\includegraphics[width=\linewidth]{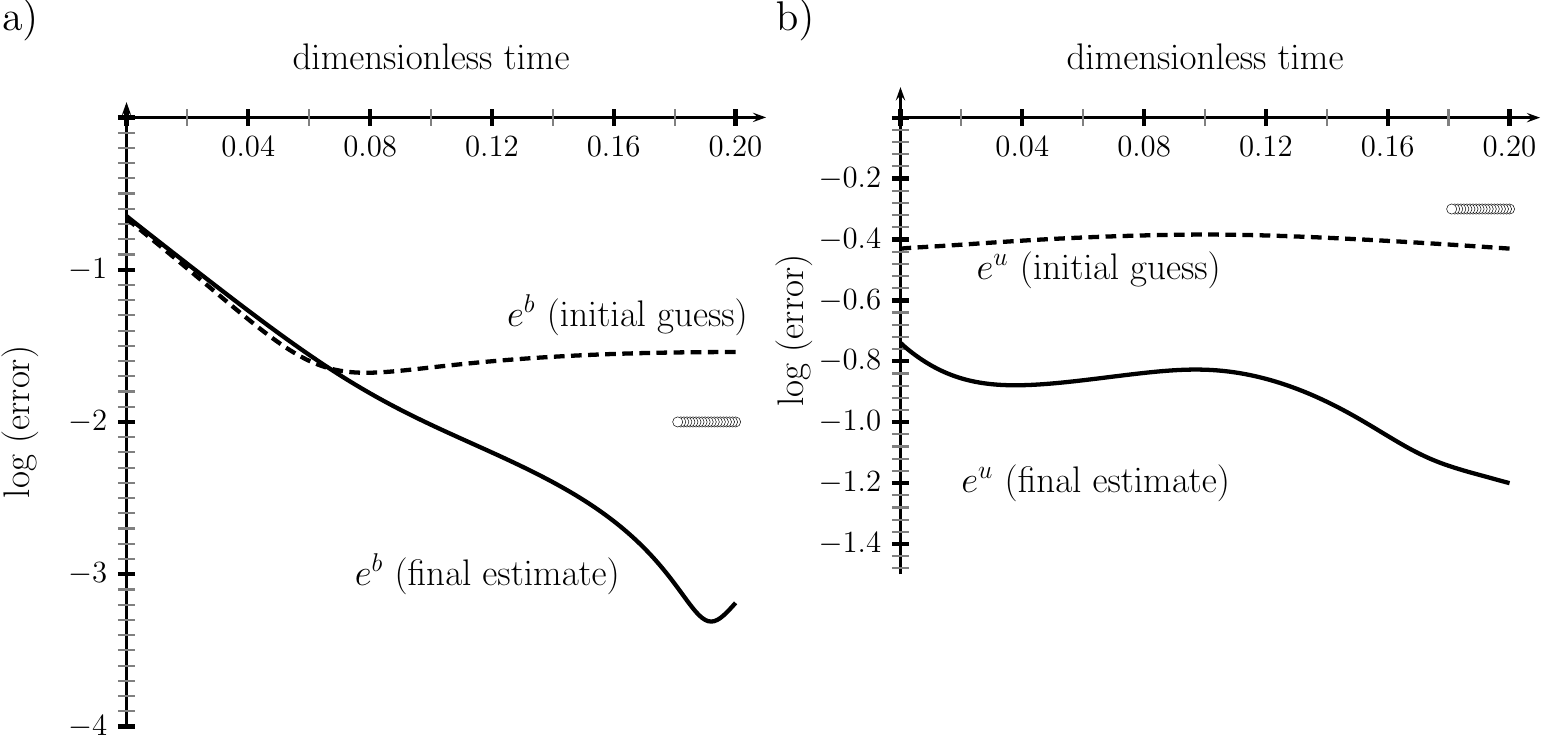}}
\caption{\label{fig:dyne_irreg} Same as Fig.~\ref{fig:dyne}, save that the $nt=20$ epochs at which 
         measurements are made are concentrated over the last $10\%$ of model 
         integration time.}
\end{figure*}

We can also assume that the temporal sampling rate is not 
constant (keeping the spatial network of observatories 
homogeneous), restricting for instance drastically the epochs at
which observations are made to the last $10$ \% of model integration 
time, the sampling rate being ideal (that is performing
observations at each model step). Not surprisingly, we
are penalized by our total ignorance of the $90$ remaining
per cent of the record. We illustrate the  
results obtained after assimilation with our now well-known
array of $\nsta=20$ stations by plotting the evolution of
the errors in $b$ and $u$ (as defined above) versus time in 
Fig.~\ref{fig:dyne_irreg}. 
Although the same amount of information ($\nsta nt = 400 $) has
been collected to produce Figs.~\ref{fig:dyne} and \ref{fig:dyne_irreg},
the uneven temporal sampling of the latter has dramatic consequences
on the improvement of the estimate of $b$. In particular, the initial error $\eb_0$
remains large. The error decreases then linearly with time until
the first measurement is made. We also observe that the minimum
$\eb$ is obtained in the middle of the observation era. The poor 
quality of the temporal sampling, coupled with the not-sufficient
spatial resolution obtained with these 20 stations, does not allow
us to reach error levels as small as the ones obtained in Fig. \ref{fig:dyne}, 
even at epochs during which observations are made. The velocity is
sensitive to a lesser extent to this effect, with velocity errors
being roughly $2$ times larger in Fig.~\ref{fig:dyne_irreg} than
in Fig.~\ref{fig:dyne}. 

\subsection{Imposing an a priori constraint on the state}
\label{sec:const}
As stated in Sect.~\ref{sec:metho}, future applications of variational
data assimilation to the geomagnetic secular variation might require
to try and impose a priori constraints on the core state. In a kinematic
framework, this is currently done in order to restrict the extent of the null
space when trying to invert for the core
flow responsible for the GSV \citep{backus1968,lemouel1984}. 

Assume for instance that we want to try and minimize the 
gradients of the velocity and magnetic fields, in a proportion
given by the ratio of their diffusivities, that is the 
magnetic Prandtl number $Pm$, at any model time. 
The associated cost function is written
\begin{equation}
J_C = \sum_{i=0}^{n} \left[\bdisc_i^T \deriv^T \deriv \bdisc_i
     +Pm \left(\udisc_i^T \deriv^T \deriv \udisc_i\right)\right],
\label{defconst}
\end{equation}
in which $\deriv$ is the derivative matrix introduced in \S\ref{sec:num}. 
The total misfit reads, according to Eq.~\eqref{defj} 
$$
J = \alpha_H J_H + \alpha_C J_C, 
$$
with $\alpha_H=1/(nt\nsta)$ as before, and $\alpha_C = \beta / [n (N-1)]$, in which 
$\beta$ is the parameter that controls the constraint to observation weight ratio. 
\begin{figure}
\centerline{\includegraphics[width=\linewidth]{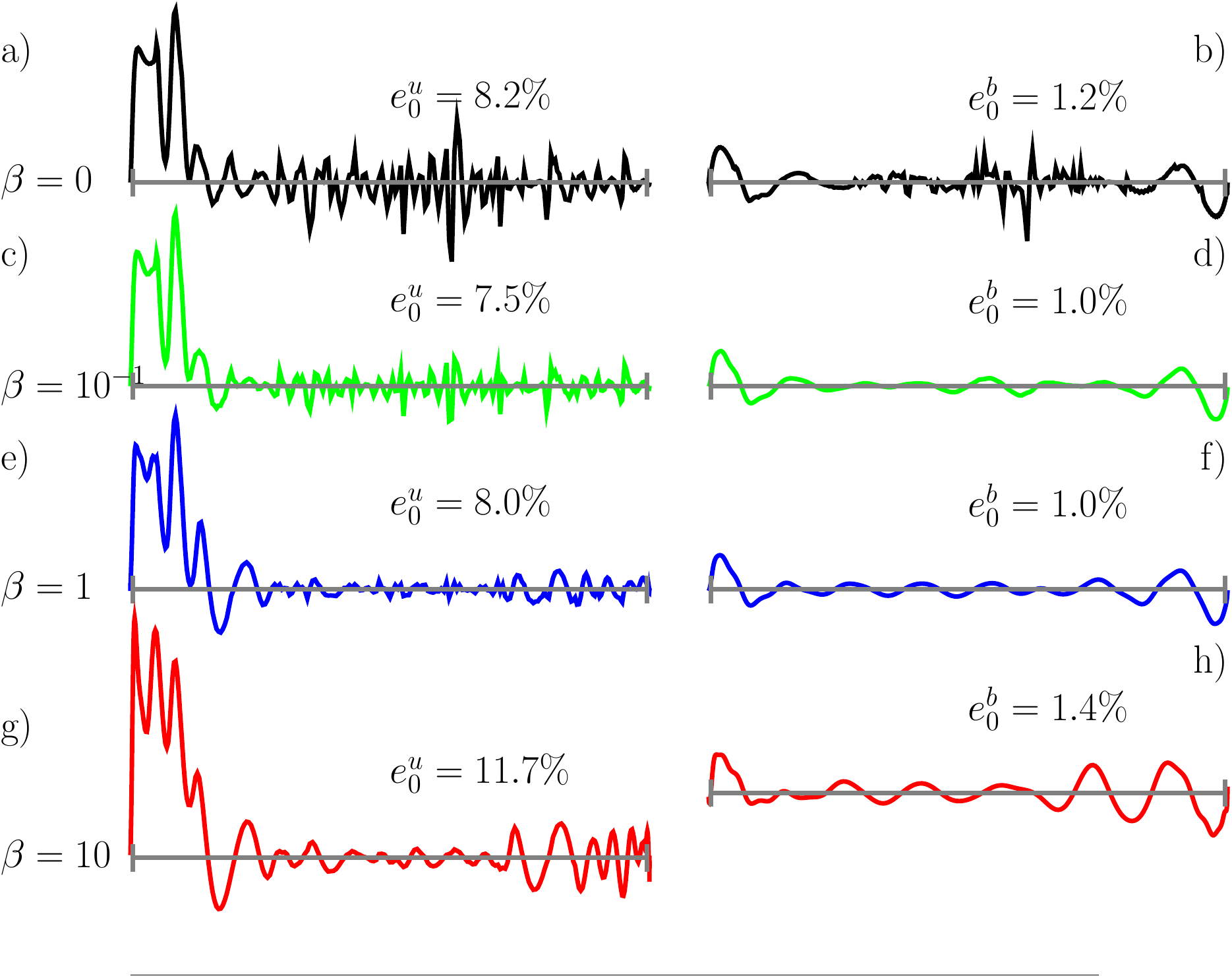}}
\caption{\label{fig:const} Influence of an a priori imposed constraint (in this case aiming at 
         reducing the gradients in the model state) on the results of variational
         assimilation. Shown are the difference fields (arbitrary scales) between
         the assimilated and true states, for the velocity field (left panel) and
         the magnetic field (right panel), at initial model time. Again, as 
         in Fig.~\ref{fig:typ}, we have made $nt=20$ measurements at $\nsta=20$ evenly 
         distributed stations. 
          $\beta$ measures the relative ratio
         of the constraint to the observations. 
         Indicated for reference are
         the $L_2$ errors corresponding to each configuration. The grey line
         is the zero line. }
\end{figure}  
Response of the assimilated model to the imposed constraint is illustrated in Fig.~\ref{fig:const}, 
using the $(nt=20,\nsta=20)$ reference case of Fig.~\ref{fig:typ}, for three increasing
values of the $\beta$ parameter: $10^{-1},1,$ and $10^1$, and showing also for reference
what happens when $\beta=0$. We show the error fields (the scale is arbitrary, but the
same for all curves) at the initial model time, for velocity (left panel) and magnetic
field (right panel). The $L_2$ errors for each field at the
end of assimilation indicate that this particular constraint can result in marginally 
better estimate of the initial state of the model, provided that the value of the 
parameter $\beta$ is kept small. For $\beta=10^{-1}$, the initial magnetic field
is much smoother than the one obtained without the constraint and makes more physical
sense (Fig.~\ref{fig:const}d). 
 The associated velocity field remains spiky, with peak to peak error amplitudes strongly
reduced in the heart of the computational domain (Fig.~\ref{fig:const}c). 
 This results in smaller errors  (reduction of about $20 \%$ for $b_0$ and $10 \%$ for $u_0$). Increasing further
the value of $\beta$ leads to a magnetic field that is too smooth (and an error
field even dominated by large-scale oscillations, see Fig.~\ref{fig:const}h), simply
because too much weight has been put on the large-scale components of $b$. The velocity
error is now also smooth (Fig.~\ref{fig:const}g), at the expense of a velocity field
 being further away from the sought solution ($\eu_0=11.7$\%), especially in the left Hartmann 
 boundary layer. 
\begin{figure}
\centerline{\includegraphics[width=\linewidth]{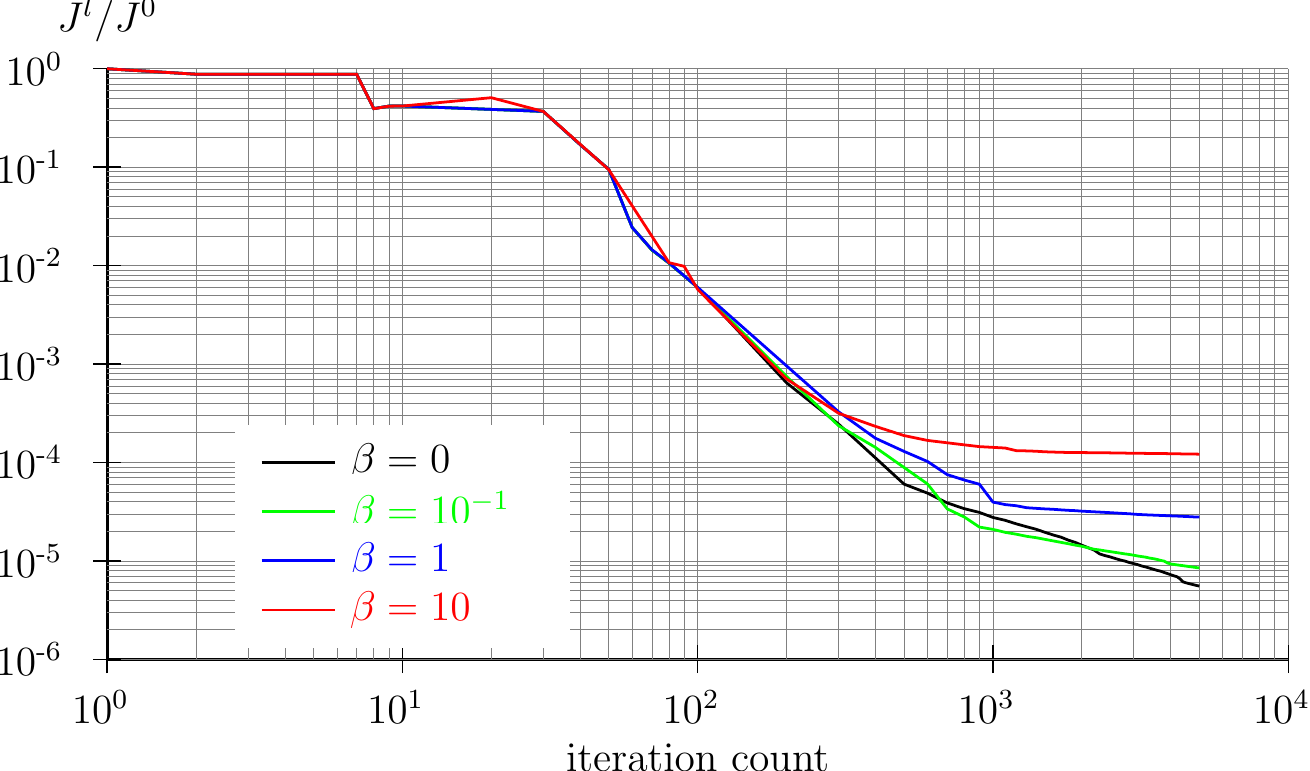}}
\caption{\label{fig:conv} Convergence behaviour for different constraint levels $\beta$. The
ratio of the current value of the misfit $J^l$ (normalized by its initial 
value $J^0$) is plotted against the
iteration count $l$. $\beta$ measures the strength of the constraint
imposed on the state relative to the observations.}
\end{figure}
 In the case of real data assimilation (as opposed to the synthetic case here, 
 the true state of which we know, and departures from which we can easily quantify),
 we do not know the true state. To get a feeling for the response of the system to 
 the imposition of an extra constraint, it is nevertheless possible to monitor
 for instance the convergence behaviour during the descent. On Fig.~\ref{fig:conv}, the
ratio of the misfit to its initial value is plotted versus the iteration number
in the conjugate gradient algorithm (log-log plot). If $\beta$ is small, the
misfit keeps on decreasing, even after 5,000 iterations (green curve). 
 On the other hand, a too strong a constraint (blue and red curves in
Fig.~\ref{fig:conv}) is not well accommodated by the model and results in a rapid flattening
of the convergence curve, showing that convergence behaviour can be used as
a proxy to assess the  efficacy of an a priori imposed constraint.  
  
Again, we have used the constraint given by Eq.~\eqref{defconst} for illustrative purposes, and 
do not claim that this specific low-pass filter is mandatory for the assimilation of GSV data. Similar types
of constraints are used to solve the kinematic inverse problem of GSV \citep{blox91}; 
see also \cite{paisetal2004} and \cite{ao2004} for recent innovative studies 
on the subject. The example developed in this section aims at showing that a formal continuity exists between the kinematic
and dynamical approaches to the GSV.  
\subsection{Convergence issues}
In most of the cases presented above, the iteration counts had reached $5,000$ before 
the cost function had decreased by $8$ orders of magnitude.  Even though
the aim of this paper is not to address specifically the matter of convergence acceleration algorithms,
 a few comments are in order, since $5,000$ is too large a number when considering
 two- or three-dimensional applications. 
 \begin{itemize}
 \item In many cases, a reduction of the initial misfit by only $4$ orders of magnitude gives
       rise to decent solutions, obtained typically in a few  
       hundreds of iterations. For example, in the case corresponding to Fig.~\ref{fig:typ}, 
       a decrease of the initial misfit by $4$ orders of magnitude is obtained after 
       $475$ iterations. The resulting error levels are already acceptable : 
       $\eu_0=12$~$10^{-2}$, $\eu_n=7.5$~$10^{-2}$, $\eb_0=1.8$~$10^{-2}$, and $\eb_n=3.0$~$10^{-4}$. 
 \item More importantly, in future applications, convergence will be sped up through
       the introduction of a background error covariance matrix $\becm$, resulting
       in an extra term \citep{icgl97}
       $$
       \frac{1}{2}[\statev_0 - \statev_b ]^T \becm^{-1} [\statev_0 - \statev_b ]
       $$
       added to the cost function (Eq.~\eqref{defj}). Here, $\statev_b$ 
       denotes the background state at model time~$0$, the definition of which 
       depends on the problem of interest.  
        In order to illustrate how this extra term can accelerate the 
       inversion process, we have performed the following assimilation experiment: we take the 
       network of virtual observatories of Fig.~\ref{fig:typ}, and 
       define the background state at model time $0$ to be zero for the velocity field (which is not
       directly measured), and the polynomial extrapolation of the $t=0$ magnetic observations made 
       at the $\nsta=20$ stations on the $N+1$ GLL grid points for the magnetic field
       (resorting to Lagrangian interpolants
       defined by the network of stations). The background error covariance
       matrix is chosen to be diagonal, without cross-covariance terms. This 
       approach enables a misfit reduction by $5$ orders of magnitude in 238 iterations, with
       the following $L_2$ error levels : 
       $\eu_0=13$~$10^{-2}$, $\eu_n=11.9$~$10^{-2}$, $\eb_0=2.6$~$10^{-5}$, and $\eb_n=2.6$~$10^{-4}$. 
       This rather crude approach is beneficial for a) the computational cost  
       and b) the estimate of the magnetic field. The recovery of the velocity is 
       not as good as it should be, because we have made no assumption at all on 
       the background velocity field. In future applications of VDA to the GSV, 
       some a priori information on the background velocity field inside the core will have
       to be introduced in the assimilation process. The exact nature of this information is beyond the scope of
       this study. 
 \end{itemize}
\section{Summary and conclusion}
\label{sec:dis}
We have laid the theoretical and technical bases necessary to 
apply variational data assimilation to the geomagnetic secular
variation, with the intent of improving the quality of the 
historical geomagnetic record. For the purpose of illustration, 
we have adapted these concepts (well established in the oceanographic 
and atmospheric communities) to a one-dimensional nonlinear 
MHD model.  Leaving aside the technical details exposed
in section \ref{sec:toy}, we can summarize our findings
and ideas for future developments as follows: 
\begin{itemize}
\item Observations of the magnetic field always have a 
      positive impact on the estimate of the invisible velocity field, 
      even if these two fields live at different length scales
      (as could be expected from the small value of the magnetic
       Prandtl number).  
\item With respect to a purely kinematic approach, having
      successive observations dynamically related by the
      model allows one to partially overcome errors due to 
      a poor spatial sampling of the magnetic field. This 
      is particularly encouraging in the prospect of 
      assimilating main geomagnetic field data, the resolution of which 
      is limited to spherical harmonic degree $14$ (say), because
      of (remanent or induced) crustal magnetization. 
\item Over the model integration time ($20$ \% of an 
      advection time), regions poorly covered
      exhibit poor recoveries of the true
      fields, since information does not have enough time
      to be transported there from well covered regions. 
      In this respect, model dynamics clearly controls 
      assimilation behaviour. 
      Concerning the true GSV, the time window we referred
      to in the introduction has a width of roughly 
      a quarter of an advective time scale. Again, this
      is rather short to circumvent the spatial 
      limitations mentioned above, if advective transport controls
      the GSV catalog. This catalog, however, could contain
      the signature of global hydromagnetic oscillations
      \citep{1966Hide,fj2003}, in which case our hope is
      that problems due to short duration and coarse spatial 
      sampling should be alleviated. This issue
      is currently under investigation in our simplified framework, 
      since the toy model presented here supports Alfv\'en waves. 
\item A priori imposed constraints (such as the low-pass filter
      of Sect.~\ref{sec:const}) can improve assimilation
      results.  They make variational data assimilation appear 
      in the formal continuity of kinematic geomagnetic inverse problems
      as addressed by the geomagnetic community over the past
      $40$ years. 
\end{itemize}
\begin{figure*}
\centerline{\includegraphics[width=\linewidth]{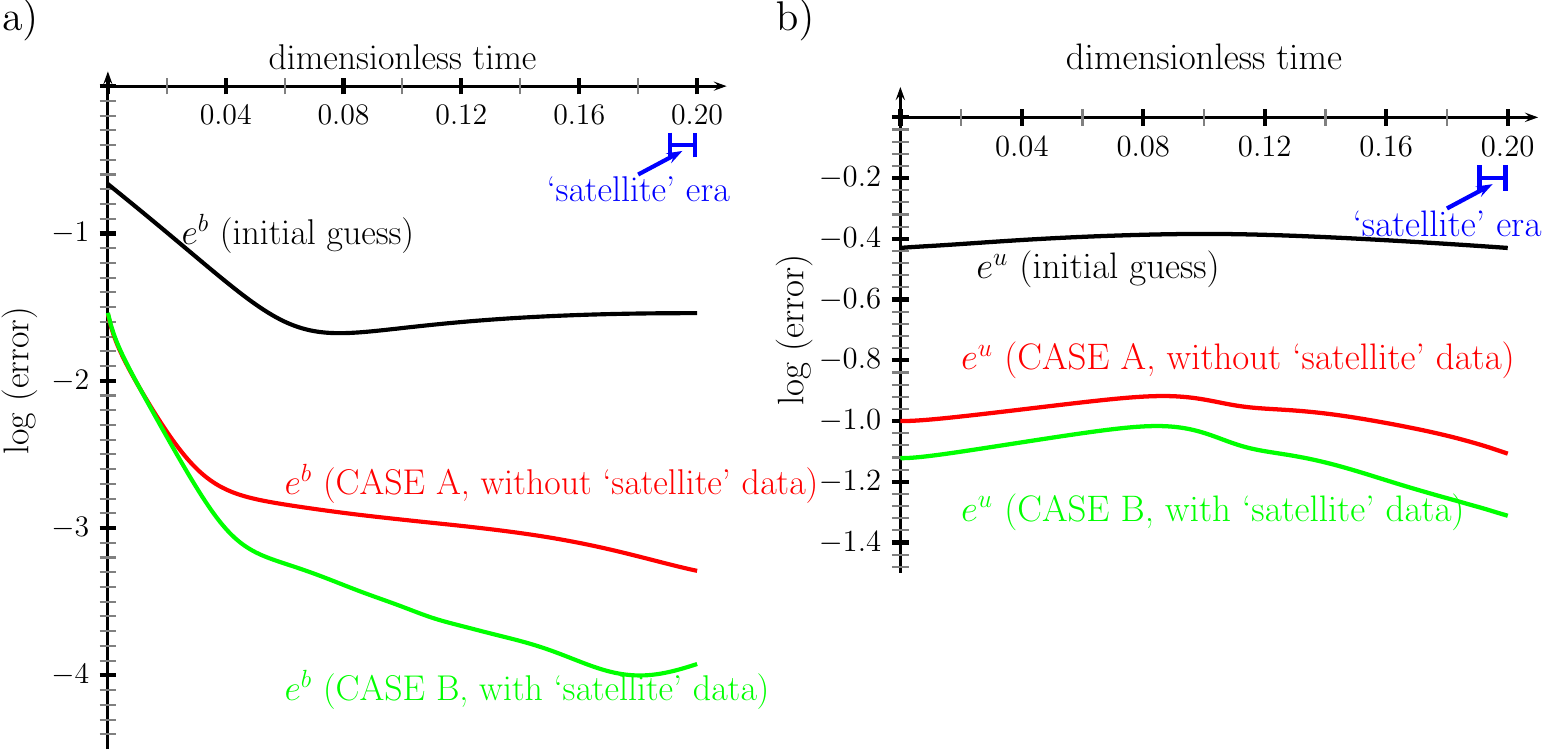}}
\caption{\label{fig:gufm} Dynamical evolution of $L_2$ errors (logarithmic value)
         for the magnetic field (a) and the fluid velocity (b).
         Black lines : errors for initial guesses. Green (red) lines : errors
         for assimilation results that do (not) incorporate the data obtained
         by a dense virtual network of magnetic stations, which aims at mimicking
         the satellite era -the blue segment on each panel-, spanning the
         last $5$ \% of model integration time.}
\end{figure*}
Finally, in order to illustrate the potential interest of applying VDA techniques
to try and improve the recent GSV record, we show in Fig.~\ref{fig:gufm}
the results of two synthetic assimilation experiments. These
are analogous to the ones described in great length in Sect.~\ref{sec:sae} 
 (same physical and numerical parameters, constraint
parameter $\beta=10^{-1}$). In both cases, 
observations are made by a network of $6$ 
evenly distributed stations during the first half of model integration
time (the logbooks era, say). The second half of the record is then produced by
a network of $15$ stations for case A (the observatory era). 
For case B, this is also
the case, save that the last $5$\% of the record are obtained
via a high-resolution network of $60$ stations. The two records 
therefore only differ in the last $5$\% of model integration
time. Case B is meant to estimate the potential impact of the 
recent satellite era on our description of the historical record. 

The evolution of the magnetic error $\eb$ backwards in time (Fig.~\ref{fig:gufm}a)
shows that the benefit due to the dense network is noticeable over 
 three quarters of model integration time, with an error 
reduction of roughly a factor of $5$. The velocity field is (as usual)
 less sensitive to the better quality of the record; still, it
responds well to it, with an average decrease of $\eu$ on the order of
$20$\%, evenly distributed over the time window. 

Even if obtained with a simplified model (bearing in particular 
    in mind that real geomagnetic observations are only available at the core  
    surface), these results are promising
and indicate that VDA should certainly be considered as the natural
way of using high-quality satellite data to refine the historical
 geomagnetic record in order to `reassimilate' \citep{tal97} pre-satellite
observations. To do so, a good initial guess is needed, which is
already available \citep{gufm}; also required is a forward model (and its adjoint) 
describing the high-frequency physics of the core. This model could either
be a full three-dimensional model of the geodynamo, or a 
two-dimensional, specific model of short-period core dynamics, based on the
assumption that this dynamics is quasi-geostrophic \citep{jault2006}. 
The latter possibility is under investigation. 

\vspace{1.cm}

{\noindent \bfseries \large Acknowledgements} 

We thank Andrew Tangborn and an anonymous referee for their very useful
comments, and \'Elisabeth Canet, Dominique Jault, Alexandra Pais,
and Philippe Cardin for stimulating discussions. AF also thanks \'Eric Beucler for sharing
his knowledge of inverse problem theory, and \'Elisabeth Canet for
her very careful reading of the manuscript. 

This work has been partially supported by a grant from the
  Agence Nationale de la Recherche ("white" research program VS-QG,
  grant reference BLAN06-2\_155316).

All graphics were produced using the freely available {\tt pstricks} and {\tt pstricks-add} packages.


\end{document}